\documentclass[twocolumn,9pt]{article} 

\usepackage[square,numbers,sort&compress,comma]{natbib}

\usepackage{amsmath}
\usepackage{amssymb}
\usepackage{caption}
\usepackage{graphicx}
\usepackage{latexsym}
\usepackage{times}
\usepackage[pagewise]{lineno}
\usepackage{hyperref}
\usepackage[version=4]{mhchem}
\usepackage{siunitx}
\usepackage{tabularx}
\usepackage{layouts}
\usepackage{comment}

\topmargin - 12pt 
\renewenvironment{abstract}%
              {
               \small
               {\bfseries \abstractname}
               \par
               \vspace{10pt}
              }

\renewcommand\abstractname{Abstract}

\newcommand{\nomenclature}
              [1]
              {
               \bgroup
               \flushleft
               \small\bf
               #1
               \par
               \egroup
              }

\renewcommand{\section}
              [1]
              {
               \bgroup
               \flushleft
               \small\bf
               \refstepcounter{section}
               \arabic{section}. #1
               \par
               \egroup
              }

\renewcommand{\subsection}
              [1]
              {
               \bgroup
               \flushleft
               \small\em
               \refstepcounter{subsection}
               \arabic{section}.
               \arabic{subsection}. #1
               \par
               \egroup
              }

\renewcommand{\subsubsection}
              [1]
              {
               \bgroup
               \flushleft
               \small\em
               \refstepcounter{subsubsection}
               \arabic{section}.
               \arabic{subsection}.
               \arabic{subsubsection}. #1
               \par
               \egroup
              }

  \newcommand{\acknowledgement}
              [1]
              {
               \bgroup
               \flushleft
               \small\bf
               #1
               \par
               \egroup
              }

  \newcommand{\sectionbib}
              [1]
              {
               \bgroup
               \flushleft
               \small\bf
               #1
               \par
               \egroup
              }
\newcommand\T{\rule{0pt}{2.6ex}}       
\newcommand\B{\rule[-1.2ex]{0pt}{0pt}} 

\usepackage[pdftex]{pdfpages} 
\usepackage{pgffor} 

\makeatletter
\AtBeginDocument{\let\LS@rot\@undefined}
\makeatother

\def\supplementfilename{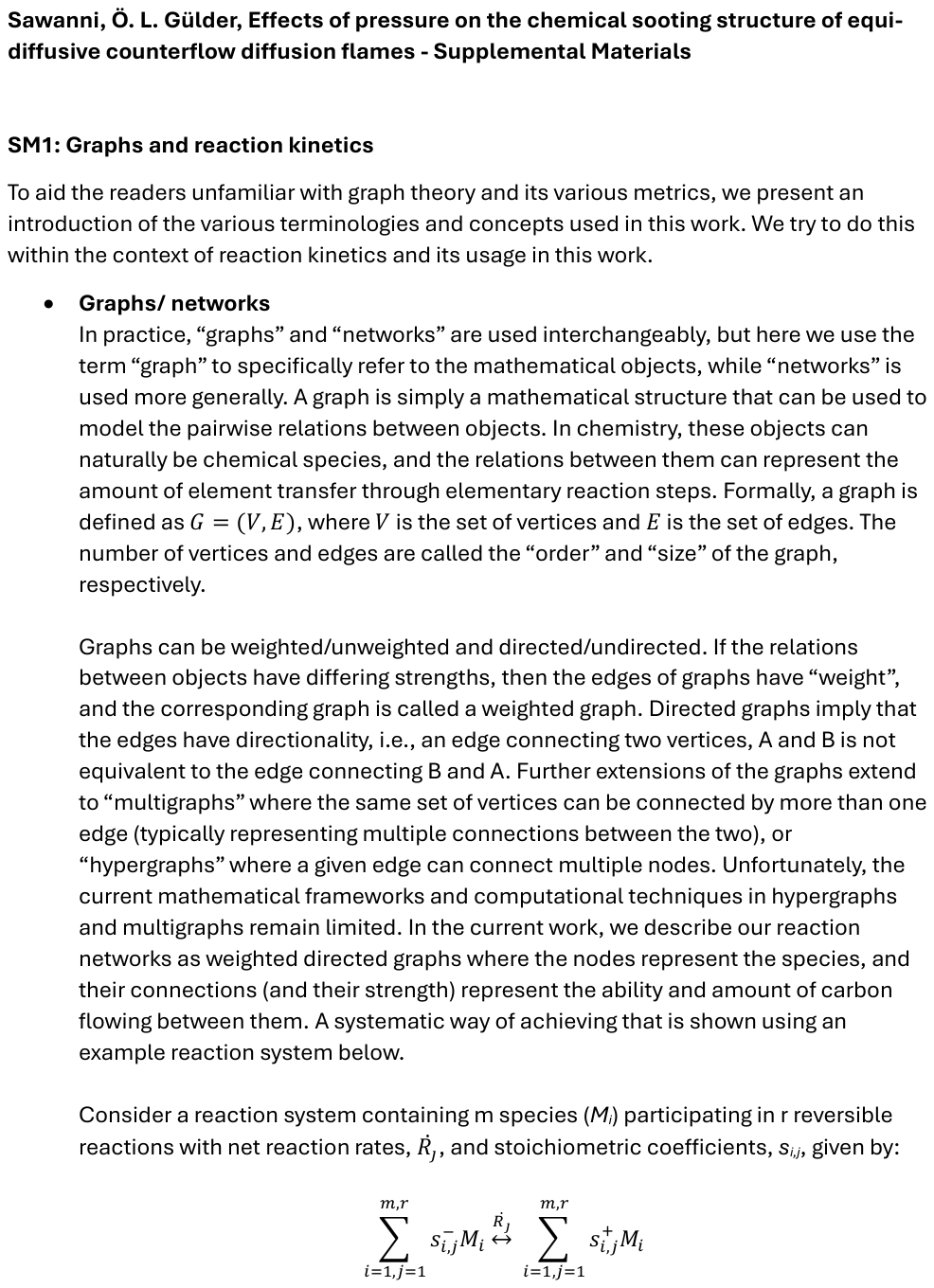}

\pdfximage{\supplementfilename}
\def\numbersupplementpages{\the\pdflastximagepages}

\newif\ifarXiv
\arXivtrue 

\begin{document}



\small
\baselineskip 10pt

\setcounter{page}{1}
\title{\LARGE \bf Effects of pressure on the chemical sooting structure of equi-diffusive counterflow diffusion flames}

\author{{\large Rajat Sawanni$^{a,*}$, \"{O}mer L. G\"{u}lder$^{a}$}\\[10pt]
        {\footnotesize \em $^a$University of Toronto Institute for Aerospace Studies, 4925 Dufferin Street, Toronto, ON M3H5T6, Canada}}

\date{}  

\twocolumn[\begin{@twocolumnfalse}
\maketitle
\rule{\textwidth}{0.5pt}
\vspace{-5pt}

\begin{abstract} 
The effects of pressure on the chemical sooting structure of equi-diffusive soot formation (SF) in a counterflow diffusion flame (CDF) are explored in a combined experimental and numerical study over pressures ranging from 1 bar to 6 bar. Experiments preserve the diffusive flame structure and carbon flux at increasing pressures and utilize measurements of soot concentrations, dispersion exponents and soot production rates. Numerical simulations are completed in OpenSMOKE++ with detailed \ce{C1-C16} chemistry, lumped PAH consideration up to \ce{C160}, sectional soot model and tracking of the C/H ratio in particulates. Network-based tools are utilized to study the organization and evolution of carbon routing pathways. Results show that for equi-diffusive flames, soot concentration increases with residence time and pressure, whereas soot production rates are influenced only by pressure. Soot C/H ratio is observed to increase with pressure using numerical and experimental methods, but numerical solutions underestimate the increase in hydrogen abstraction reactions. The soot-forming network undergoes a percolation-like organization of its pathways before soot inception. The network then continues to grow by adding connections through its influential nodes. Acetylene is identified as a highly influential node in the carbon transfer graph, with its influence increasing with pressure.
\end{abstract}

\vspace{10pt}

{\bf Novelty and significance statement}

\vspace{10pt}
This work advances understanding of the pressure effects on soot formation chemistry using a combined experimental and numerical work in CDFs. Utilization of multi-wavelength absorption-emission diagnostics for the measurement of soot concentration, production rates and maturity provides a rich dataset to the limited body of high-pressure sooting CDF data. Moreover, the study advances soot chemistry analysis by casting it as a chemical reaction network (CRN) and extending pathway analysis beyond a single dominant pathway to a graph‑theoretic framework that elucidates soot‑reaction topology.

\vspace{5pt}
\parbox{1.0\textwidth}{\footnotesize {\em Keywords:} Soot formation; high pressure; networks}
\rule{\textwidth}{0.5pt}
*Corresponding author.
\vspace{5pt}
\end{@twocolumnfalse}] 

\section{Introduction\label{sec:introduction}}

Soot formation in hydrocarbon flames remains a critical area of combustion research due to the complex coupling of chemical kinetics and transport processes during particle inception and growth \cite{wang2019b, karatas2012}. Beyond the complexity of its formation processes, soot plays a dominant role in the composition of pollutant emissions, climate radiative forcing \cite{bond2013}, and the degradation of practical systems \cite{bozzano2002}, while simultaneously providing a promising pathway for the controlled synthesis of carbon nanomaterials \cite{kammler2001}. Pressure is a key control parameter in this problem, as it alters the soot chemistry by changing collisional rates, radical lifetimes and third‑body efficiencies \cite{karatas2012}, ultimately affecting the growth routes and chemical pathways that control the sequestration of carbon to soot. Disentangling the purely chemical effects of pressure, however, remains non-trivial in the presence of concurrent changes in transport processes \cite{thomson2005, liu2006, abdelgadir2017}.

Counterflow diffusion flames (CDFs), specifically soot-forming (SF) CDFs \cite{kang1997}, provide a unique and well-controlled experimental and modeling framework for isolating the chemical effects of soot production from oxidation and other transport processes. This framework is enabled by their purely diffusional nature \cite{tsuji1982}, quasi-1D flow field \cite{zhang2023}, independent control of strain rates \cite{bohm2001}, and minimal heat loss to the burner walls \cite{figura2012}. One of the key effects of increasing pressure is the reduction of diffusive transport \cite{figura2014, karatas2012, abdelgadir2017, gleason2019, xue2018, axelbaum1988, linan1974}. Unlike the typically studied buoyancy-dominated co-flow diffusion flame, which experiences thinning flames as a result of decreasing diffusivity at increasing pressures \cite{karatas2012, abdelgadir2017}, CDFs can be designed to study the effects of pressure in flames where the diffusive thickness remains constant with increasing pressures, i.e. an \emph{equi-diffusive} flame structure \cite{xue2018, axelbaum1988, sun1996, fotache1995}. This is achieved by maintaining density-weighted strain rates as pressure increases. It is not completely surprising that this experimental design also maintains similar carbon flux into the flame at increasing pressures \cite{gleason2019, xue2018, axelbaum1988}, albeit at the cost of changing residence times. 

Previous works on high-pressure SF CDFs have exploited their control to study soot formation in constant strain rates and density-weighted strain rate arrangements. Naturally, disparate pressure-scaling of soot concentrations for these experimental designs has been reported in the literature \cite{amin2023, gleason2024, xue2018, sawanni2024}. Building on the works of Gleason et al. \cite{gleason2023, gleason2024}, we conducted an experimental campaign with both experimental designs to reconcile the pressure scaling in CDFs \cite{sawanni2024}. It was noted that despite the disparate pressure scaling exponents of soot concentrations, mean volumetric soot production rates ($\dot{\omega_{s}}'''$) remained consistent across the two experimental designs. Moreover, simple chemical markers, i.e., acetylene and pyrene concentrations, were found to collapse soot production rates across different experimental designs and peak flame temperatures. Interestingly, independent measurements of sooting limits by Du et al. \cite{du1998}, Sung et al. \cite{sung1998}, and Sarnacki and Chelliah \cite{sarnacki2018} also observed an increase in soot production rate with first-order dependence on acetylene concentrations. Similar scaling for polycyclic aromatic hydrocarbon (PAH) fluorescence was noted by B{\"o}hm and Lacas \cite{bohm2000} near sooting limits. In an allied study by the authors \cite{sawanni2026}, we investigated the structure and chemistry of soot formation when pressure was varied while maintaining constant residence times. Single dominant pathways were extracted at each location that maximized carbon flux from fuel species (\ce{C2H4}) to soot. The study reiterated the prominent roles of PAH adsorption reactions and presented a unique distribution of global pathways across the spatial domain. Unlike a co-flow diffusion flame \cite{zhou2021}, soot inception was observed to be led by acetylene addition to larger PAHs, while smaller PAHs dominated its subsequent growth. Under increasing pressures, the planar-PAH growth route was strengthened, and the dominance of the acenaphthylene and pyrene-mediated routes spanned the entire soot production domain.

The present study inverts the previous protocol and investigates the effects of pressure when carbon flux to the flame is held constant across pressure, while allowing residence time to vary. Soot aging, or soot maturity, provides an additional measure characterizing the carbon-to-hydrogen ratio of soot and offering further insight into soot nucleation and growth processes \cite{cepeda2025, blanquart2009, celnik2009, raj2010, saggese2015}. Soot maturity not only serves as a measure of soot age within the flame, enabling identification of potential nucleation zones, but also governs the optical properties of the generated soot particles. Recent studies suggest that incipient soot particles have smaller absorption functions ($E(m)$) \cite{bescond2016}, which depend on wavelength through the dispersion exponent, $\beta$, such that $E(m) \propto \lambda^{-\beta}$ \cite{michelsen2021}. Improved characterization of soot particles enhances measurement accuracy \cite{yon2021} and aids in constraining their environmental impact \cite{bond2013} as well as their potential use in nanomaterial synthesis \cite{kelesidis2017, kammler2001}.

This study further extends network-based analysis from single dominant pathways to a backbone-based network analysis approach. Although network-based methods have been successfully applied to a wide variety of dynamical systems \cite{albert2002, newman2003, li2017}, their use within the combustion and soot formation community remains comparatively limited \cite{mutlay2015}. Graph-based representations of soot and reaction chemistry are not new and have been employed for mechanism generation \cite{gao2016rmg}, reduction \cite{lu2005, sun2010}, and reaction pathway analysis \cite{sun2010, gao2016, gao2019, zhou2021}. As post-processing tools, pathway analysis methods range from simple rate-of-production analyses involving immediate neighbors (single-generation graphs) to global source-to-sink pathway analyses (multi-generational graphs) \cite{gao2019}. The first step in these approaches is the construction of a graph in which species are represented as nodes and reactions define their connections. Depending on the objective, edges may represent stoichiometric relationships \cite{lu2005}, reaction fluxes \cite{sun2010}, or, in some cases, elemental fluxes between species \cite{zhou2021}. Dominant pathways can then be identified using graph-based algorithms such as Dijkstra’s shortest path method \cite{dijkstra1959}.

Despite their success in providing skeletal mechanisms and kinetic insights for large, intractable systems \cite{zhou2021}, graph-based tools have largely been restricted to pathway identification and mechanism reduction, with relatively few studies using them to gain deeper kinetic insights \cite{zhou2021, sawanni2026}. Beyond pathway extraction, network theory offers a range of metrics that characterize reaction network topology and provide information on hub centralities, network growth, and structural transitions. Mutlay and Restrepo \cite{mutlay2015} examined such properties in pyrolysis and combustion networks and showed that their time-dependent growth exhibits features of an infinite-order phase transition, along with a unique correlation between percolation thresholds and the electron distributions of the reactants.

This work addresses three pertinent questions. First, how does the soot network organize its dominant pathways that deliver carbon to soot at different positions in the counterflow flame? Second, when pressure is increased, how do the sooting structures evolve, and how do these network measures and pathways reorganize when carbon flux is held constant? Third, how do network‑level reorganizations correlate with measured soot properties, and can acetylene and pyrene compositions be interpreted as potential descriptors of the pathway fluxes? To achieve these objectives, measurements of soot concentration and dispersion exponents are combined with numerical solutions to provide valuable insights into the CDF sooting processes. We then construct a graph-based representation of the CRN at each location and extract the soot subgraph, which contains species and connections that capture most of the carbon flux to soot. Dominant routes from all species to soot are considered and several graph metrics are utilized to present a unique insight into the global sooting process at increasing pressure.

\section{Methodology\label{sec:methods}}\vspace*{-0.2cm}
\subsection{Experimental methods}
A counterflow burner, designed and housed in the high-pressure chamber at UTIAS \cite{sawanni2024, thomson2005}, was used for this study. The experimental facility and related P\&ID were discussed in detail in our previous work \cite{sawanni2024, sawanni2026}, while some salient details are briefly introduced here. The burner consists of opposing central contracting nozzles with an exit diameter of \SI{10}{mm} and a contraction ratio of $16$. The bottom nozzle is supplied with a mixture of nitrogen and ethylene, while the top nozzle is supplied with a mixture of nitrogen and oxygen. Two surrounding co-flowing nozzles shroud the central flow and maintain flame stability at high pressures. Experiments were designed to maintain stoichiometric mixture fraction ($z_{\textrm{st}} = 0.226$) and a density-weighted strain rate with increasing pressures, as shown in Table \ref{tab:exp_cond}. The fuel and oxygen mass fractions in their respective streams are maintained at $y_{\mathrm{f}} = y_{\mathrm{ox}} = 0.35$, which create a highly sooting flame with peak temperatures ranging from $T_{\mathrm{ma }} = 2327-2628$ K. The cases with similar letter labels (A, B, C) in Table \ref{tab:exp_cond} maintain similar density-weighted strain rates, while the following numeric label indicates the pressure. To avoid flame instability at high pressure and low strain rates, the density-weighted strain rates are maintained in steps. The flow is simultaneously simulated using a 2-D buoyantReactingFoam solution to determine the exit boundary conditions for the quasi-1D code. The extracted centreline velocity boundary conditions and their gradients are also detailed in Table \ref{tab:exp_cond}.

\begin{table}[!t]
	\centering
	\caption{Test conditions, $U_{\mathrm{f}},~U_{\mathrm{ox}}$: centreline velocity at the fuel and oxidizer nozzle exit, respectively, $V_{\mathrm{r,f}},~V_{\mathrm{r,ox}}$: radial velocity at the fuel and oxidizer nozzle exit, respectively, $p$: pressure, $T_{\mathrm{max}}$: computed peak flame temperature}
	{\footnotesize
		\resizebox{\columnwidth}{!}{
			\begin{tabular}{|c|c|c|c|c|c|c|c|c|c|}
				\hline 
				Case & $U_{\mathrm{f}}$ & $U_{\mathrm{ox}}$ & $\frac{\mathrm{d}V_{\mathrm{r, f}}}{\mathrm{d}r}$ & $\frac{\mathrm{d}V_{\mathrm{r, ox}}}{\mathrm{d}r}$ & $p$ & $T_{\mathrm{max}}$\T \\
				& m/s & m/s & s\textsuperscript{-1} &  s\textsuperscript{-1} & bar & K \\
				\hline
				A1   & 0.265 & 0.246 & 31.1 & 33.2 & 1 & 2327\T \\
				A2  & 0.128 & 0.128 & 16.5 & 15.2 & 2 & 2500 \\
				\hline
				B2   & 0.232 & 0.221 & 27.1 & 27.8 & 2 & 2464\T \\
				B3  & 0.150 & 0.153 & 19.4 & 17.1 & 3 & 2548 \\
				\hline
				C4   & 0.213 & 0.207 & 23.8 & 23.1 & 4 & 2570\T \\
				  C5  & 0.165 & 0.174 & 20.5 & 16.9 & 5 & 2604 \\
				C6  & 0.133 & 0.149 & 17.8 & 13.0 & 6 & 2628\B \\
				\hline
			\end{tabular}
	}}\vspace*{-12pt}
	\label{tab:exp_cond}
\end{table}

A modulated absorption-emission technique is employed with a diffuse light source to simultaneously evaluate soot concentrations, dispersion exponent and soot temperature. The diagnostic layout is described in detail in the previous work \cite{sawanni2024, sawanni2025, sawanni2026}, with the inclusion of extinction and emission at multiple wavelengths ($650, ~700, ~750, ~800, ~ 850, ~900~ \textrm{nm}$). The collection optic was calibrated for emission measurements using a NIST traceable calibration lamp housed within an integrating sphere (StellarNet UIS-LS 6"). The emission signal was corrected for signal trapping as reported in \cite{sawanni2025}. Using the reconstructed radial emission field, the temperature is evaluated using single-color self-absorption corrected absolute radiance ($J_{\lambda}$) and extinction coefficient ($k_{\mathrm{ext,\lambda}}$). 
\begin{align}
	T = \left(-\frac{k_b\lambda}{hc}\mathrm{ln}\left(\frac{\lambda^5J_{\lambda}}{2hc^2k_{\mathrm{ext, \lambda}}}\right)\right)^{-1}
\end{align}
where $\lambda, h,c,k_b$ are the wavelength, Planck's constant, speed of light and Boltzmann constant, respectively. Following temperature evaluation, the dispersion exponent can be evaluated using the relation, 
\begin{align}
	a + \beta\mathrm{ln}\lambda = \mathrm{ln}\left(\frac{B_{\lambda}(T)}{J_{\lambda}\lambda}\right)  \label{eq:beta}
\end{align}
where $B_{\lambda}(T)$ is the black-body radiance at wavelength, $\lambda$ and temperature, $T$. Finally, soot volume fraction, $f_{\mathrm{v}}$, is calculated using the measured extinction coefficient at \SI{650}{nm} assuming a constant scattering albedo, $\alpha_{\lambda}=0.1$, and dispersion exponent corrected soot absorption function, $E(m_{\lambda},\beta)$ \cite{bescond2016}.
\begin{align}
	f_v = \frac{k_{\mathrm{ext, \lambda}}\lambda}{6\pi E(m_{\lambda}, \beta)(1+\alpha_{\lambda})}
\end{align}
The errors associated with the measurement of emission and extinction profiles are taken from our previous work \cite{sawanni2025}, and propagated to currently reported measurements using error propagation analysis (see supplementary section).

\subsection{Numerical methods}
The numerical results are obtained with the CounterFlowDiffusion solver of the OpenSMOKE++ solver suite \cite{cuoci2013} and follow a similar methodology as  in \cite{sawanni2024, sawanni2026}. The high-temperature CRECK mechanism is utilized \cite{saggese2015}, which couples detailed isomer resolved gas-phase chemistry up to 4 ring PAHs, lumped approach for PAHs ranging from \ce{C20}-\ce{C160} and a discrete sectional model for soot particulate treatment. Considerations for different H/C ratios are also included, which are then translated to numerical dispersion exponents using the relations of Michelsen et al. \cite{michelsen2021} ($\mathrm{H}/\mathrm{C} = 0.39\beta - 0.27$). The validation and modeling accuracy of the implemented methods were compared in a recent work \cite{sawanni2026} and are added in the supplementary section for completeness. The evaluation of soot production rates utilizes a combination of experimental and numerical methods and follows our previous work \cite{sawanni2024}. Measured SVF is translated to local soot mass fraction $Y_{\mathrm{s}} = f_{\mathrm{v}}\rho_{\mathrm{s}}/\rho$ where $\rho_{\mathrm{s}} = \SI{1800}{kg/m^{3}}$ is the soot density and $\rho$ is the local gas density computed using OpenSMOKE++ suite. A flux-based analysis of the soot advection equation is subsequently used to compute the soot production rates using simulated velocity fields.

\subsection{Network analysis}
\begin{figure}
	\centering
	\includegraphics[width=\columnwidth]{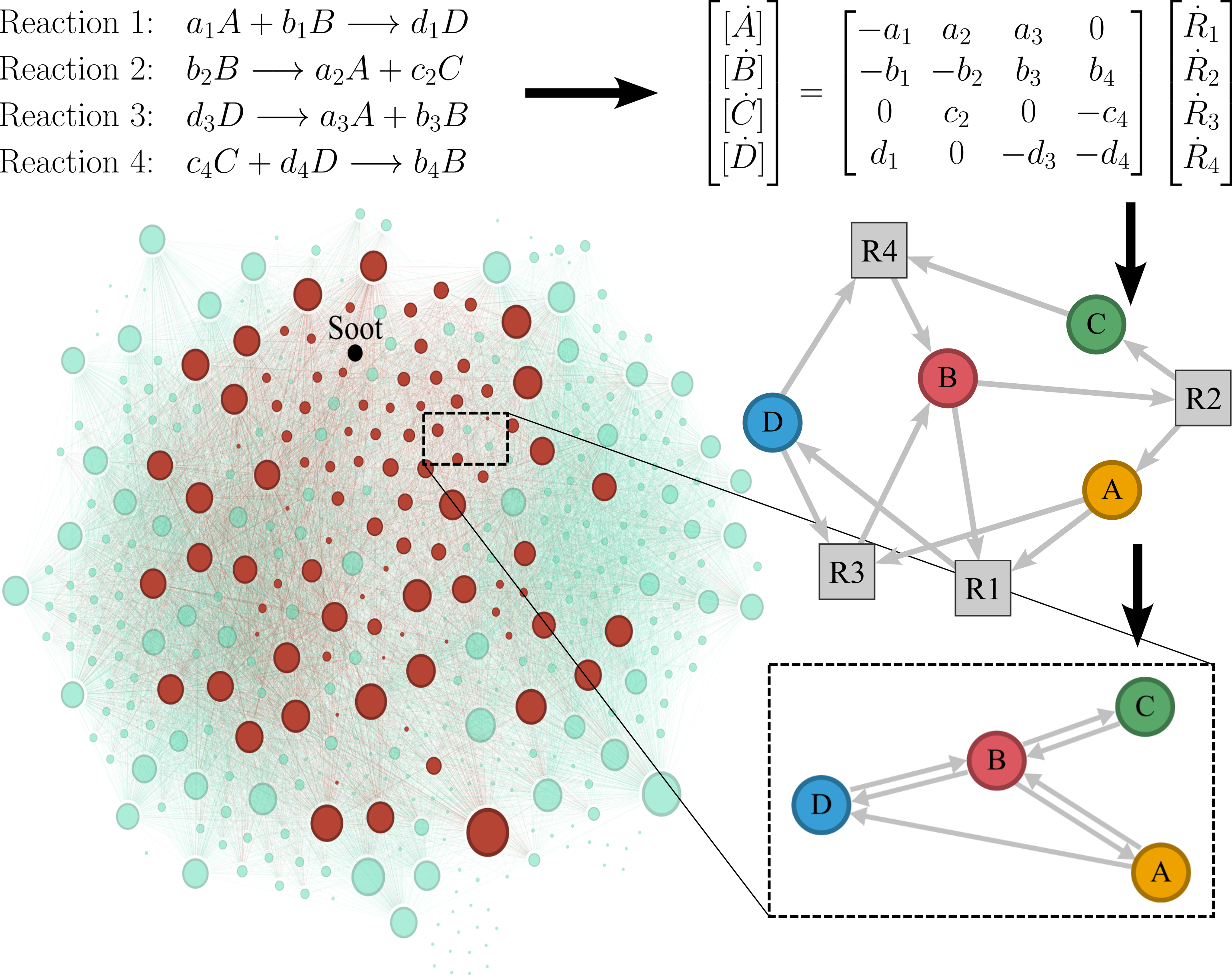}
	\caption{The procedure for converting simulated data at a given location to carbon flux graph from a sample set of reactions with known reaction rates. Also shown is the overall flux graph and extracted soot subgraphs with nodes in blue and maroon, respectively}
	\label{fig:graphs}\vspace*{-12pt}
\end{figure}
The current work extends the analysis of global pathways from generated networks in previous works \cite{sawanni2026, zhou2021} to a graph-based network analysis framework. A graph is hereby simply an abstraction of the CRN where each node represents a specie, and edges denote connections between specific pairs of species. Since network based analysis is used herein as a post-processing tool, the simulation results are used as input for the graph generation. The overview of graph generation process is shown in Fig. \ref{fig:graphs}. NetworkX \cite{hagberg2008} python library is used to generate directed graphs and analyze various metrics. Note that a directed graph ensures that the connection (here, carbon flux) from species A to B is not equivalent to that from B to A. For any given case and location within the flame, a directed carbon flux graph is then constructed where each node denotes a species and its edges denote the total carbon flux through all reactions involving them. 
\begin{align}
	A_{\mathrm{C}, i \to j} &= \sum_{r} \mathrm{max}\left(0,~B_{\mathrm{C}, r, i \to j}\dot{R}_{r}\right)
\end{align}
where the contribution of each individual reaction, $r$, is denoted through its net reaction rate, $\dot{R}_{r}$ and the number of carbon atoms transferred from species $i$ to species $j$, $B_{\mathrm{C}, r, i \to j}$ in that reaction.
\begin{align}
	B_{\mathrm{C}, r, i \to j} &=
	\begin{cases}
		n_{\mathrm{C}, r, j}\frac{n_{\mathrm{C}, r, i}}{n_{\mathrm{C}, r}}, & \text{if } \nu_{r, j}\nu_{r, i} < 0 \\
		0, &\text{otherwise}
	\end{cases}
\end{align}
where $n_{\mathrm{C},r,i}$ and $n_{\mathrm{C},r,j}$ denote the total number of carbon atoms transferred out of species $i$ and into species $j$, respectively, in reaction $r$; $n_{\mathrm{C},r}$ is the total number of carbon atoms transferred in reaction $r$; and $\nu_{r,i}$ and $\nu_{r,j}$ are the stoichiometric coefficients of species $i$ and $j$ in reaction $r$. To maintain directionality, net negative reaction rates are ignored for connections between $i\to j$ but are included in the flux contribution from $j\to i$.

To ensure that edge weights and path distances have a clear physical interpretation, we construct a new graph ($F$) from the original graph ($G$) by preserving all nodes and renormalizing the edge weights
\begin{align}
	F_{\mathrm{C}, i \to j} = -\log\left(\frac{A_{\mathrm{C}, i \to j}}{\sum_{k, k \neq j} A_{\mathrm{C}, k \to j}}\right).
\end{align}
Let $p_{ij,k} = (i, \ldots, j)_k$ denote the $k^{\mathrm{th}}$ ordered path connecting species $i$ to species $j$. Under this renormalization, the path distance, $d_{i \to j, p_{ij,k}}= \sum_{l=1}^{N_p-1} F_{\mathrm{C}, p_{ij,k}(l) \to p_{ij,k}(l+1)}$, represents the fractional carbon flux delivered to $j$ along $p_{ij,k}$, where $N_p$ is the path length. Essentially, if the total carbon flux into species $j$ is unity, the flux through the path is given by $\exp\!\left(-d_{i \to j, p_{ij,k}}\right)$. This formulation enables the use of standard graph-theoretic routing algorithms to extract physically meaningful measures of soot-forming topology. We quantify the connectivity of soot (species $s$) in the network using its throughput, $\mathcal{T}_{\mathrm{s}}$, defined as the mean maximum carbon flux that any species can contribute to soot:
\begin{align}
	\mathcal{T}_{\mathrm{s}} = \exp\left(-\frac{1}{N}\sum_{i=1}^{N_G} d_{i \to s, p_{is,1}}\right),
\end{align}
where $d_{i \to s, p_{is,1}}$ denotes the shortest path from species $i$ to soot in graph $F$, whose size is $N_G$.

Although the normalized carbon flux graph $F$ captures carbon transport among all species at a given flame location, its usefulness for soot-kinetics analysis is limited by numerous irrelevant nodes and edges. To focus the analysis and maintain computational tractability, we extract a subgraph consisting only of nodes and edges that contribute meaningfully to carbon transport toward soot. Specifically, we retain edges that belong to paths contributing at least 0.001\% of the total carbon flux to soot. The total fractional carbon flux to soot passing through a node then defines its node influence, $I_i$:
\begin{align}
	I_i = \sum_{k=1}^{N_k} \exp\left(-d_{i \to s, p_{is,k}}\right),
	\label{eq:influence}
\end{align}
where the summation includes only $N_k$ paths that satisfy the flux threshold.

\begin{figure*}[!t]
	\centering
	\includegraphics[width=\textwidth]{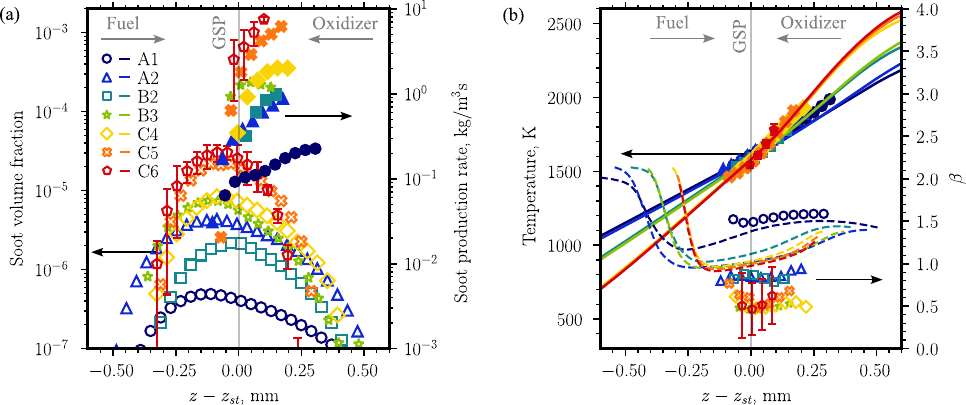}
	\caption{(a) Influence of pressure on the sooting structure of equi-diffusive soot formation CDFs. Open symbols show experimental soot volume fractions (SVF), filled symbols show evaluated soot production rates, (b) Experimental (symbols) and numerical (lines) profiles of temperature and dispersion exponents.}
	\label{fig:hydrodynamics}\vspace*{-12pt}
\end{figure*}

\section{Results\label{subsec:results}}
\subsection{Flame aerodynamics and sooting structure}

We examine the effects of pressure on flame sooting structure and aerodynamics using measurements of soot volume fraction (SVF) and soot production rate (SPR) in Fig.~\ref{fig:hydrodynamics}(a), and temperature and dispersion exponents, together with their simulated profiles, in Fig.~\ref{fig:hydrodynamics}(b). Fuel and oxidizer issue from opposing nozzles and are axially convected toward the gas stagnation plane (GSP). The high-temperature flame sheet is located on the oxidizer side of the GSP, consistent with the soot-formation (SF) CDF flame structure. The temperature profiles overlap in discrete steps with increasing pressure, in accordance with the experimental design, which maintains similar diffusive structures across conditions. While peak flame temperatures increase with pressure at a given density-weighted strain rate (see Table~\ref{tab:exp_cond}), the temperature profiles within the soot-forming regions remain unchanged. For these equi-diffusive flames, however, residence times decrease with increasing pressure. Consequently, the observed changes in soot concentration result from the combined effects of pressure and residence time.

A comparison of cases A2 and B2 isolates the effect of strain rate at constant pressure. In this case, increasing strain rate leads to a reduction in diffusive thickness, as observed in the temperature profiles. Remarkably, peak soot production rates remain approximately unchanged with changing strain rate, indicating that the observed differences in soot concentration are largely driven by residence-time effects. The influence of pressure is then clearly reflected in the scaling of soot production rates, which appears largely unaffected by residence time variations\cite{sawanni2024}. Comparing SVF and SPR profiles show that while soot production peaks near high-temperature regions, soot concentration continues to increase in a temperature-decaying environment. In contrast to gaseous precursors, which pyrolyze and flow toward the GSP from the fuel nozzle, soot formed on the oxidizer side of the GSP travels toward the fuel nozzle and stagnates at the particle stagnation plane (PSP), located slightly downstream of the GSP. This results in a sooting structure in which aged soot particles grow in the presence of smaller PAHs under decreasing temperatures before stagnating at the PSP \cite{sawanni2026}.

The soot maturity profile inferred from the dispersion exponents further complements this picture and is shown in Fig.~\ref{fig:hydrodynamics}(b). Since flow converges toward the GSP from both sides, the dispersion exponent profiles indicate that soot continues to grow and mature after inception near the high-temperature flame, progressing toward the GSP in a temperature-decaying environment. This behavior is corroborated by higher $\beta$, indicative of younger soot \cite{bescond2016, yon2021, liu2018}, near the oxidizer side of the GSP, followed by its gradual decay toward the PSP. Simulations also predict a sharp increase in dispersion exponents near the PSP. Such an increase across the PSP has been observed by Gleason and Gomez \cite{gleason2018, gleason2019, gleason2021} and attributed to a low-temperature nucleation regime. An alternative explanation was previously discussed \cite{sawanni2026} suggesting $C/H$ change due to size-dependent particle diffusion across the PSP. Increasing pressure and residence time decrease soot H/C ratios in the growth region, as expected from longer residence times and enhanced dehydrogenation and soot growth at elevated pressures. Simulations reproduce the qualitative pressure-dependent trends observed experimentally; however, the predicted dispersion exponents are consistently higher than measured values. This discrepancy may indicate that current mechanisms underpredict the rates of key dehydrogenation and hydrogen-abstraction reactions \cite{frenklach1987, hwang2001}, which play a central role in PAH growth and soot inception. A direct comparison is further complicated by the reliance of experimental dispersion exponents on the empirical formulation of Michelsen et al.~\cite{michelsen2021}, whose assumed optical properties may not fully capture soot formed in high-pressure CDFs.

\subsection{Topology of soot formation CRNs}

One of the most informative descriptors of network topology is the vertex degree distribution, where the degree of a node is defined as the number of edges connected to it. This distribution is commonly examined using a cumulative distribution function (CDF), which represents the probability of finding nodes with degree greater than $k$ \cite{mutlay2015}. This CDF, interestingly, has a unique structure for specific properties of the graph. For example, networks with degree distributions of the form $p(k) \sim k^{-\alpha}$ are commonly referred to as scale-free, as certain moments of the distribution diverge depending on the value of $\alpha$. Mechanisms based on preferential attachment, often described as ``rich-get-richer'' growth models, have been proposed as a means of generating scale-free network topology \cite{albert2002}. The presence of such topology implies a dominant role of highly connected hub nodes, which can strongly influence transport and information flow within the network. 
\begin{figure}[t!]
	\centering
	\includegraphics[width=\columnwidth]{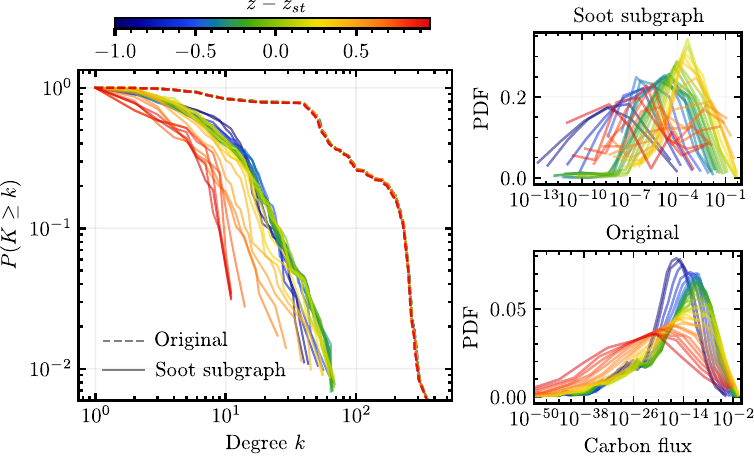}
	\caption{The cumulative degree distribution of the original and extracted soot subgraph at one of the representative locations from A1 flame. The distribution of edge weights (carbon flux) on these graphs is also shown in the inset.}
	\label{fig:degrees}\vspace*{-12pt}
\end{figure}
While scale-free networks have been reported across a range of dynamical systems, recent studies suggest that they are less prevalent in natural and engineered networks than previously assumed \cite{broido2019}. Motivated by these observations, the degree distributions of the chemical reaction networks (CRNs) considered here were systematically analyzed to assess the presence of scale-free characteristics \cite{broido2019}. The analyzed networks were found to transition between exponential and weakest scale-free characteristics depending on the flame location. This is noted to be true for both, the original as well as the soot subgraph. The degree distributions of the original graph and the extracted soot subgraph are shown in Fig.~\ref{fig:degrees} for the A1 flame at different locations. For locations where soot production rates are low (red shaded curves), the degree distribution is more consistent with an exponential form, $p(k) \sim \tau \exp\bigl(\tau (k_{\mathrm{min}} - k)\bigr)$, where $\tau$ is a rate parameter and $k_{\mathrm{min}}$ is the cutoff degree. Exponential degree distributions are commonly observed in dynamically evolving networks such as CRNs \cite{mutlay2015, deng2011}, particularly when growth proceeds through non-equilibrium or time-dependent processes that limit preferential attachment. At these locations, the carbon flux exhibits an approximately log-normal distribution with a relatively large variance, indicating that most nodes carry small amounts of carbon flux. Further downstream in the soot evolution region (blue shaded curves in Fig.~\ref{fig:degrees}), the total carbon flux through the network increases, and the flux distribution shifts toward higher values while becoming narrower. The extracted soot subgraph follows a similar trend and develops a heavier tail in its degree distribution. This behavior indicates the emergence of hub-like nodes with degrees larger than those expected in purely random, exponentially growing networks. 

\begin{figure}[h!]
	\centering
	\includegraphics[width=\columnwidth]{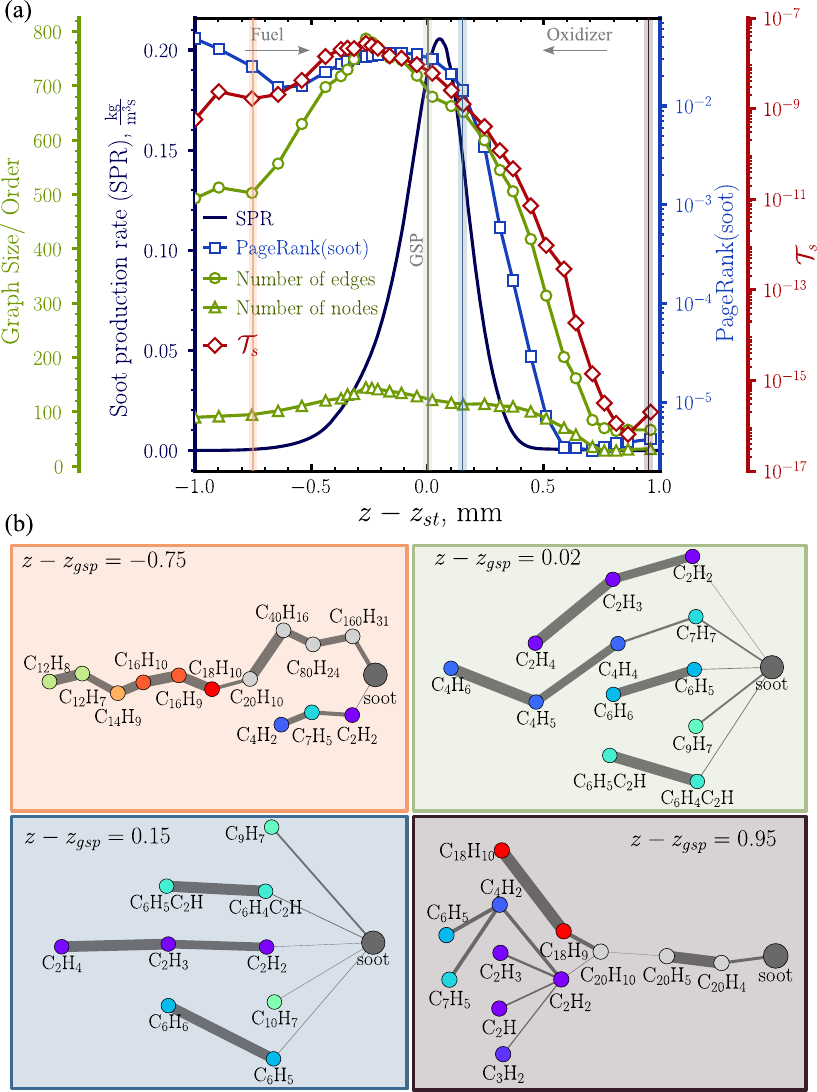}
	\caption{(a) The evolution of soot PageRank, throughput ($\mathcal{T}_s$), subgraph size, and order across the spatial domain of case A1 flame. Soot production rate profiles are overlaid for comparison. (b) Backbone graphs showing the carbon transfer from influential nodes to soot at various representative locations. Edge thickness scales with the fraction of carbon flux relative to the total flux received by the downstream node.}
	\label{fig:networkmetrics}\vspace*{-15pt}
\end{figure}

Figure~\ref{fig:networkmetrics} shows the evolution of network-based metrics for the A1 flame, overlaid with soot production rate profiles. In addition to the size of the extracted soot subgraph, the PageRank metric and soot throughput ($\mathcal{T}_s$) is also shown. Note that the PageRank of soot node and its throughput are calculated in the original network, instead of the soot subgraph. Originally developed by Google \cite{bianchini2005}, PageRank is a probability distribution that represents the likelihood that a random walker traversing the graph will visit a given node. Interestingly, PageRank was recently utilized as a centrality measure for ranking species and efficient reduction of combustion chemistry \cite{wang2025}. When teleportation is neglected, as in this work, PageRank corresponds to the stationary distribution of a Markovian random walk on the network. In the present context, this metric provides a measure of carbon's long-time visitation probability of soot at a given flame location. A corollary of the soot PageRank is its throughput, $\mathcal{T}_s$, which quantifies the reachability of soot node within the graph by estimating the geometric mean of maximum carbon flux that a specie contributes to the overall soot growth at that location.

Interestingly, all the network measures defined here, i.e., the graph size, soot PageRank, and soot throughput, exhibit a steady increase prior to the onset of detectable soot inception or measurable increases in soot production rate (Fig.~\ref{fig:networkmetrics}(a), as soot progresses from right to left along the spatial coordinate). This behavior suggests that the soot-forming reaction network assembles and strengthens carbon-delivery pathways before observable soot nucleation begins. Such behavior is reminiscent of percolation-like assembly processes reported in other dynamically evolving networks \cite{deng2011}. Once established, the organized chemical reaction network persists throughout the primary soot production zone before gradually weakening as the flow approaches PSP.

The backbone representation of the soot subgraph and its evolution along the flame is shown in Fig.~\ref{fig:networkmetrics}(b) and corresponds to the union of dominant carbon-carrying pathways identified through highly influential nodes. Prior to the network transition, carbon delivery to soot relies on a single narrow chain of links connecting large PAHs and acetylene. These links expand as the soot graph becomes more vibrant in its immediate connections hubs. Soot accumulates mass not only through a linear chain of growing PAHs, but through multiple parallel pathways connecting large parts of the graph and contributing modestly to soot growth. The contributions from smaller PAHs are noted to expand at this stage. At the full scale of the graph, these hubs siphon carbon flux through various species and starting nodes before contracting again to a single dominant chain of linear PAH growth near the particle stagnation plane.

\begin{figure}[!t]
	\centering
	\includegraphics[width=\columnwidth]{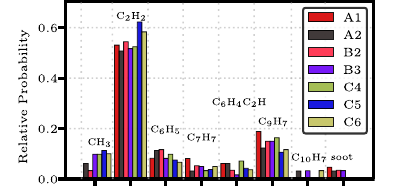}
    \caption{The probability of attachment of new nodes to the listed hub nodes across the various cases. The analysis is done in the regions of increasing graph size as soot evolves through a temperature-decaying environment.}
	\label{fig:percolation}\vspace*{-12pt}
\end{figure}

We now analyze the region where the soot subgraph grows by finding the key species and pathways through which carbon transfer to soot increases. Considering that the carbon flow from any species to soot happens through one of the immediate connections of soot (hubs), we determine the probability of attachment of new species to any one of the existing hubs or the formation of new hubs during the growth process. Fig. \ref{fig:percolation} shows these relative probabilities for the collection of hub nodes. The formation of new hub connectors is shown through soot as the 'hub' node.

Acetylene clearly emerges as the dominant hub for soot network growth, consistent with the observation of first-order dependence of sooting limits on acetylene concentration \cite{sarnacki2018} and soot production rate \cite{sawanni2024}. Its attachment probability increases modestly with pressure. Although planar PAHs and their hydrogen-abstracted radicals influence soot production \cite{sawanni2024}, their attachment does not significantly drive soot subgraph growth; however, they are among the first species to attach directly to soot during the network transition. The key contributors to the growth of soot networks are noted to be smaller radicalized PAHs like phenyl, ethynyl-benzene, benzyl and indenyl. The probabilities of attachment to all these hubs slightly increase with increasing pressure, while the formation of new hubs decreases.

\subsection{Pathways of soot production}

\begin{figure*}[h!]
	\centering
	\includegraphics[width=\textwidth]{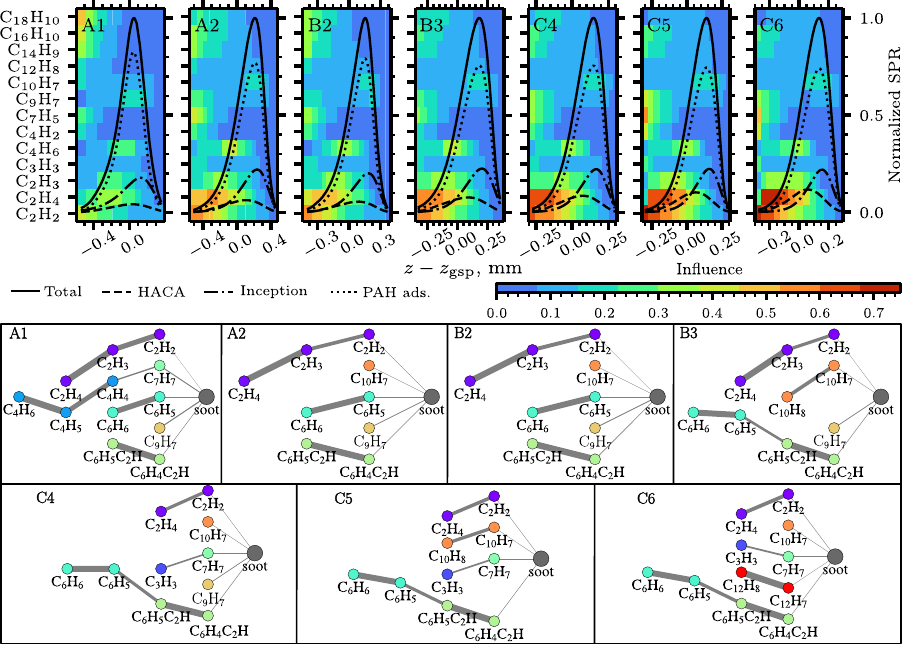}
	\caption{\emph{Top:} Lines denote the soot production rate and contributions from HACA, inception, and PAH adsorption (right y axis), while colors indicate species influence at each location, ordered by increasing molar mass (left y axis); \emph{Bottom:} Backbone graphs depict carbon transfer from influential nodes to soot at the location of peak SPR for each case. Edge thickness scales with the fraction of carbon flux relative to the total flux received by the downstream node.}
	\label{fig:influence}\vspace*{-12pt}
\end{figure*}

As opposed to the global pathway analysis from one species (usually, fuel species) to soot, the influence metric defined in Eq. \ref{eq:influence} can be used to analyze the contributions of multiple pathways starting from any species and going to soot. The influence of a given species then shows the fractional carbon flux that passes through the species and goes to soot. Figure~\ref{fig:influence} identifies critical species and shows the evolution of their influence as soot develops through the production zone for various cases. Also shown are the relative contributions of soot inception, hydrogen-abstraction carbon-addition (HACA), and PAH adsorption pathways. The relative contributions of these sub-mechanisms are observed to remain largely unchanged with pressure, a trend also reported for high-pressure CDFs at constant strain rates \cite{sawanni2026}. Starting from the high-temperature region on the right, acetylene emerges as the first influential species directing carbon flux to soot. With increasing pressure, both the magnitude and spatial extent of acetylene’s influence increase. As the soot network grows, PAH radicals such as indenyl (\ce{C9H7}) and naphthyl (\ce{C10H7}), as well as smaller species including ethylene (\ce{C2H4}), vinyl (\ce{C2H3}), and butadiene (\ce{C4H6}), exhibit increasing influence. The growing influence of smaller species is attributed to their strengthened connections through the acetylene hub. While the influence of the indenyl radical decreases with pressure, the influence of smaller fuel species and radicals increases, which may be attributed to the suppression of RSR-dominant pathways due to enhanced collisional stabilization at elevated pressures \cite{liu2006}. Near the particle stagnation plane, the influence of pyrene and planar PAHs increases; however, at higher pressures, even this contribution decreases in favor of smaller-species and acetylene-mediated pathways.

On the bottom panel, Figure~\ref{fig:influence} shows the effect of pressure on the soot backbone structure at peak soot production locations. While soot formation regions near the flame sheet and particle stagnation plane are influenced by pressure, the backbone structure at peak soot production remains largely unchanged. The relative contributions from various hubs, however, is noted to change modestly, in accordance with the observed trends in influence. The contributions from indenyl radical reduces in favor of naphthyl and acenaphthyl radicals. Other changes to the backbone structure are minimal, indicating the formation of an increasingly saturated network in which additional soot growth proceeds mainly through strengthening existing pathways rather than through substantial structural reorganization.

\section{Conclusions\label{sec:conclusions}}
With the overarching goal of delineating pressure effects on soot chemistry in counterflow diffusion flames, pertinent questions on the organization and dominance of carbon routing pathways were analyzed. Combining experimental soot concentrations, production rates, and dispersion exponents with numerical simulations reveals that evaluated soot production rates are largely insensitive to strain rate; but differences in soot concentration arise from effects of residence-time and pressure. Soot maturity (C/H ratios) is noted to increase with increasing pressure and residence time, and is qualitatively captured by the numerical code. Network analysis reveals that the soot-forming network assembles itself and strengthens the carbon-forming routes to deliver carbon into soot, even before the start of soot nucleation processes, exhibiting a percolation like transition. During the transition, the soot graph continues to grow via addition of nodes, mostly through acetylene-mediated routes. Indenyl and smaller PAHs contribute substantially in this region, with increasing influence at high pressures, perhaps due to the increasing influence of acetylene hub in the overall carbon transfer routes. 

\acknowledgement{CRediT authorship contribution statement}

{\bf Rajat Sawanni}: Writing – original draft, Methodology, Investigation, Formal analysis, Data curation, Conceptualization  {\bf \"{O}mer L. G\"{u}lder}: Writing – review \& editing, Methodology, Supervision, Resources, Funding acquisition, Conceptualization. 

\acknowledgement{Declaration of competing interest}

The authors declare that they have no known competing financial interests or personal relationships that could have appeared to influence the work reported in this paper.

\acknowledgement{Acknowledgments}
The authors thank the Natural Sciences and Engineering Research Council of Canada for the grant (RGPIN-2023-04914) supporting this research work. RS is also thankful to Dr. Amitesh Roy for the valuable discussions that helped in shaping this work. 
\footnotesize
\baselineskip 9pt

\clearpage
\thispagestyle{empty}
\bibliographystyle{proci}
\bibliography{references}

\ifarXiv


\section*{Supplementary material}

**Sawanni, Ö. L. Gülder, Effects of pressure on the chemical sooting structure of equi-diffusive counterflow diffusion flames - Supplemental Materials**

## SM1: Graphs and reaction kinetics

To aid the readers unfamiliar with graph theory and its various metrics, we present an introduction of the various terminologies and concepts used in this work. We try to do this within the context of reaction kinetics and its usage in this work.

- **Graphs/ networks**
  In practice, “graphs” and “networks” are used interchangeably, but here we use the term “graph” to specifically refer to the mathematical objects, while “networks” is used more generally. A graph is simply a mathematical structure that can be used to model the pairwise relations between objects. In chemistry, these objects can naturally be chemical species, and the relations between them can represent the amount of element transfer through elementary reaction steps. Formally, a graph is defined as $G = (V, E)$, where $V$ is the set of vertices and $E$ is the set of edges. The number of vertices and edges are called the “order” and “size” of the graph, respectively.

  Graphs can be weighted/unweighted and directed/undirected. If the relations between objects have differing strengths, then the edges of graphs have “weight”, and the corresponding graph is called a weighted graph. Directed graphs imply that the edges have directionality, i.e., an edge connecting two vertices, A and B is not equivalent to the edge connecting B and A. Further extensions of the graphs extend to “multigraphs” where the same set of vertices can be connected by more than one edge (typically representing multiple connections between the two), or “hypergraphs” where a given edge can connect multiple nodes. Unfortunately, the current mathematical frameworks and computational techniques in hypergraphs and multigraphs remain limited. In the current work, we describe our reaction networks as weighted directed graphs where the nodes represent the species, and their connections (and their strength) represent the ability and amount of carbon flowing between them. A systematic way of achieving that is shown using an example reaction system below.

  Consider a reaction system containing m species ($M_i$) participating in r reversible reactions with net reaction rates, $\dot{R}_J$, and stoichiometric coefficients, $s_{i,j}$, given by:

$$\sum_{i=1,j=1}^{m,r} s_{i,j}^{-} M_i \overset{\dot{R}_J}{\leftrightarrow} \sum_{i=1,j=1}^{m,r} s_{i,j}^{+} M_i$$

R1: $a_1A + b_1B \leftrightarrow d_1D$
R2: $b_2B \leftrightarrow c_2C + a_2A$
R3: $a_3A + d_3D \leftrightarrow b_3B$
R4: $c_4C + d_4D \leftrightarrow b_4B$
...... and so on.

A representation of the mass action law using linear algebra then shows:

$$\frac{d\mathbf{M}}{dt} = \mathbf{S}.\mathbf{R}$$

$$d\begin{bmatrix}[A]\\ [B]\\ [C]\\ [D]\\ \vdots\end{bmatrix}\Big/ dt = \begin{bmatrix}-a_1 & a_2 & a_3 & 0 & \cdots\\ -b_1 & -b_2 & b_3 & b_4 & \cdots\\ 0 & c_2 & 0 & -c_4 & \cdots\\ d_1 & 0 & -d_3 & -d_4 & \cdots\\ \vdots & \vdots & \vdots & \vdots & \ddots\end{bmatrix}\begin{bmatrix}\dot{R}_1\\ \dot{R}_2\\ \dot{R}_3\\ \dot{R}_4\\ \vdots\end{bmatrix}$$

Where **M** is the array of species concentrations, **S** is the stoichiometric coefficient matrix, and **R** is the array of net reaction rates. A natural representation of the above reaction system can be made using a bipartite graph (Temkin et al., 2020), where we have two sets of vertices defining the species and reactions. Each reaction node connects all its associated species, where the directionality and weights represent the production or consumption rate of species through that reaction. The bipartite representation of a small part of the above chemical reaction network is shown as:

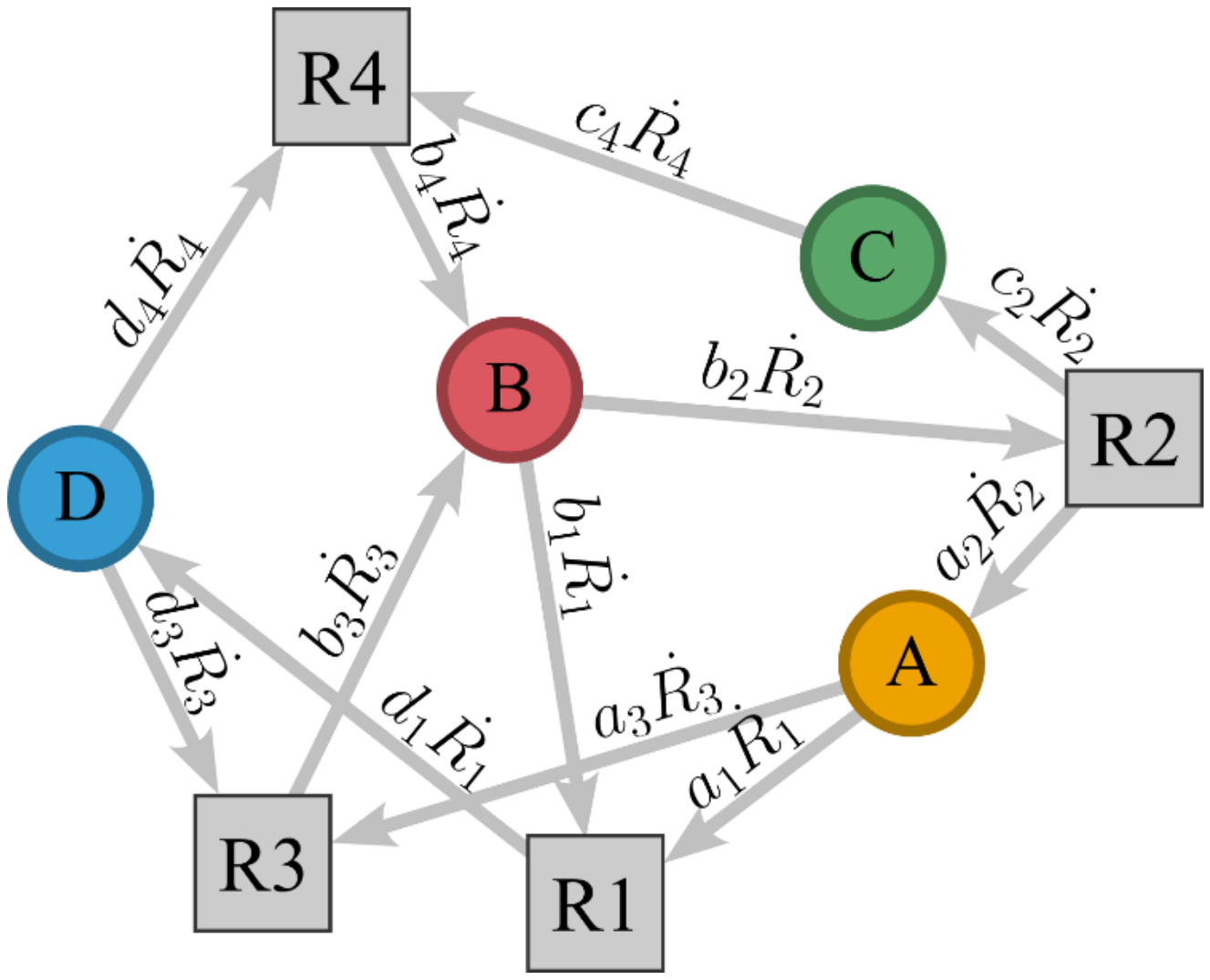


*Figure S1: Bipartite representation of the mass action law*

Since bipartite directed weighted graphs have limited practical computational techniques, we usually employ the species graphs, which are unipartite projections of the above bipartite graphs. The projection of this bipartite graph for an undirected, unweighted graph is straightforward, and is shown below:

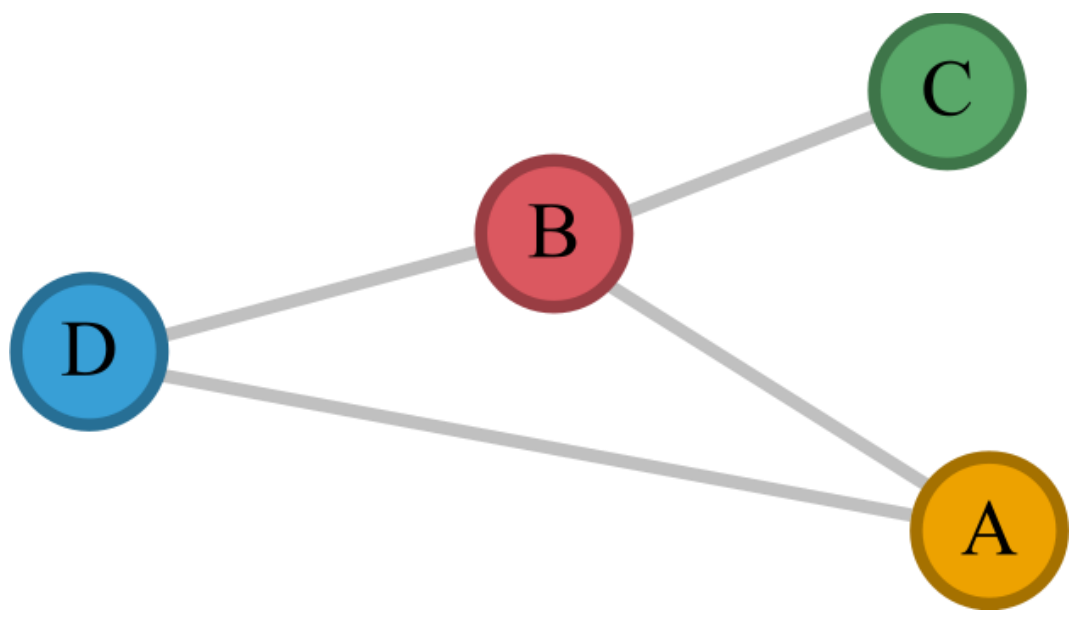


*Figure S2: Unipartite projection of the reaction network with only species nodes and edges denoting their connections through reactions. Essentially, if two species are on opposite sides of a reaction, an edge is created between them. This is an undirected and unweighted graph.*

The above species projection connects two species, if they exist on the reactant and product side of a reaction. This undirected graph represents a reaction network where species interconvert freely. A physical interpretation that can be easily made by looking at this graph is that A cannot convert to C without passing through B, making B a hub intermediate.

A more detailed projection of the bipartite graph can be obtained if we discuss the elemental flow within the system. This makes natural sense for a species projection, as connections between species nodes can naturally be used to represent the amount and direction of elemental flow within the system. If the elemental (hereby carbon) composition of each specie, $M_i$, is given by $n_{c,i}$, the elemental carbon transfer for each reaction can be given by:

R1: $\mathrm{a_1A + b_1B \leftrightarrow d_1D}$ $\quad n_{c,A}a_1 + n_{c,B}b_1 = n_{c,D}d_1$

R2: $\mathrm{b_2B \leftrightarrow c_2C + a_2A}$ $\quad n_{c,B}b_2 = n_{c,C}c_2 + n_{c,A}a_2$

R3: $\mathrm{a_3A + d_3D \leftrightarrow b_3B}$ $\quad n_{c,D}d_3 + n_{c,A}a_3 = n_{c,B}b_3$

R4: $\mathrm{c_4C + d_4D \leftrightarrow b_4B}$ $\quad n_{c,C}c_4 + n_{c,D}d_4 = n_{c,B}b_4$

...... and so on.

For each reaction, the overall carbon transfer from a given specie to another can be written in a similar way and is shown below using a weighted directed graph.

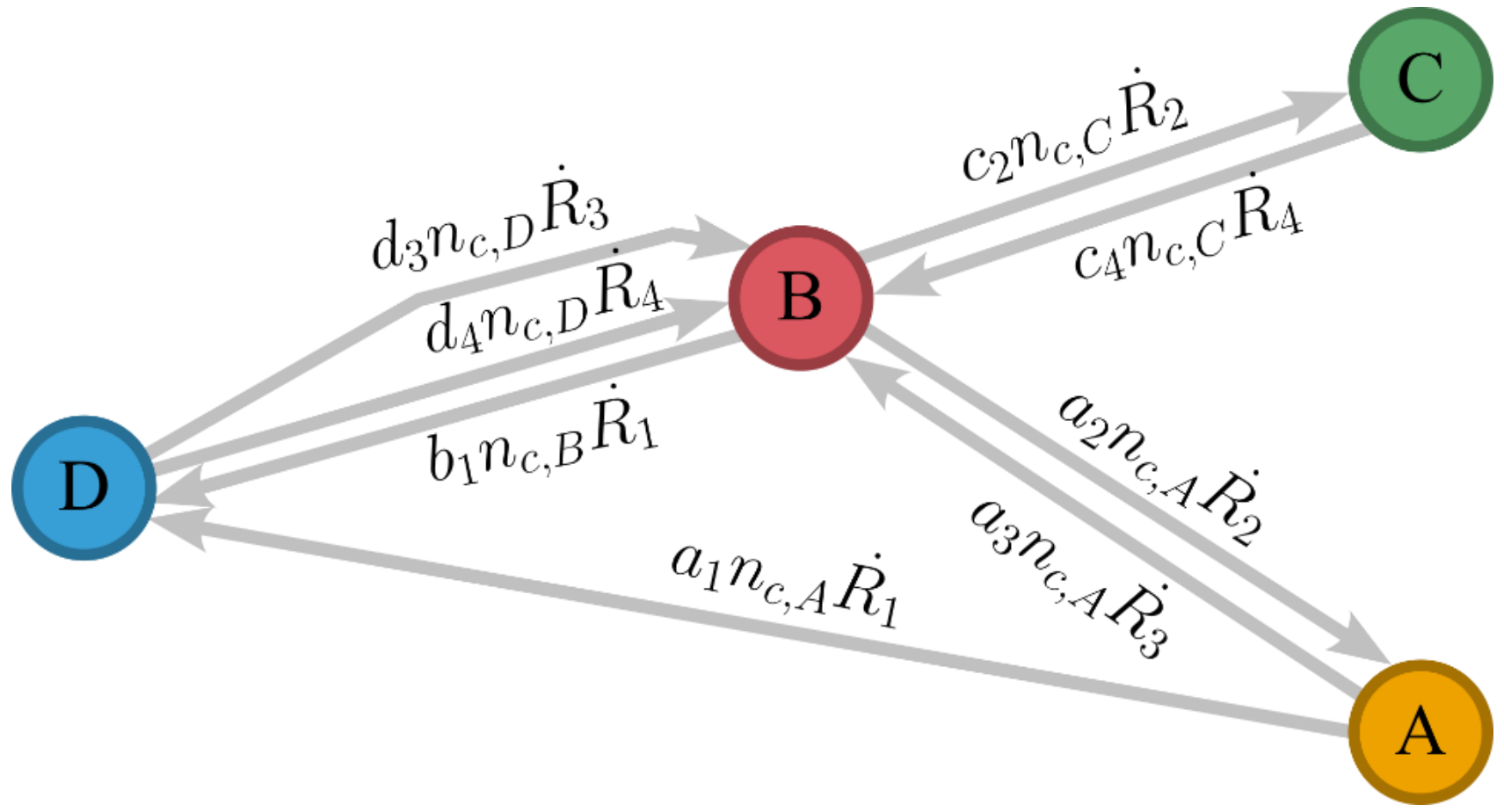


*Figure S3: Directed weighted graph representation showing the carbon flux through all reactions between them. Each edge denotes a connection and carbon flux through a specific reaction.*

Note that for each pair of species, there could be multiple edges denoting the carbon transfer from each reaction. It is also possible that one or more of these arrows are reversed, depending on whether the net reaction rates are positive or negative. The net resultant weighted directed graph would then be the sum of all the positive contributions for an ordered pair of species. This implies that if all reactions except R3 have a net positive reaction rate, our directed weighted graph would be represented as:

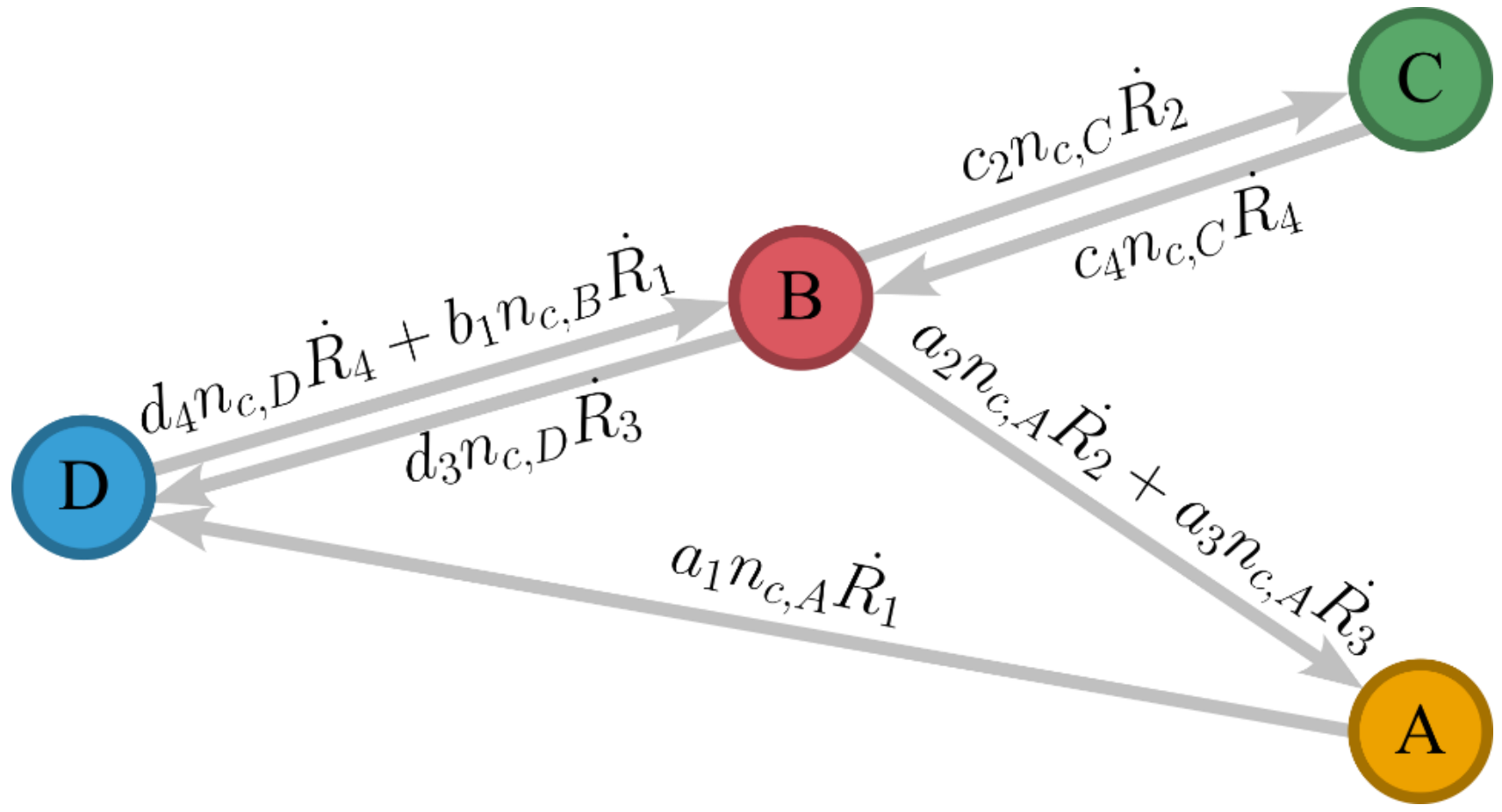


*Figure S4: Directed weighted graph representation showing the carbon flux through all reactions between them. Each edge denotes a connection and carbon flux through all reactions between the pair of species.*

The final graph now has all the important features that are critical for identifying dominant species, pathways and reaction mechanisms within the context of carbon flow. Since the basis of its creation is still the mass action law of reaction kinetics, the graph and its topology essentially highlight the structure of carbon flow within the chemical reaction network. The relationship between consumed and produced carbon for this graph can now be denoted using the linear relation:

$$\begin{bmatrix} n_{c,A}[A]_p \\ n_{c,A}[B]_p \\ n_{c,A}[C]_p \\ n_{c,A}[D]_p \end{bmatrix} = \begin{bmatrix} 0 & a_2 n_{c,A}\dot{R}_2 + a_3 n_{c,A}\dot{R}_3 & 0 & 0 \\ 0 & 0 & c_4 n_{c,C}\dot{R}_4 & d_4 n_{c,D}\dot{R}_4 + b_1 n_{c,B}\dot{R}_1 \\ 0 & c_2 n_{c,C}\dot{R}_2 & 0 & 0 \\ a_1 n_{c,A}\dot{R}_1 & d_3 n_{c,D}\dot{R}_3 & 0 & 0 \end{bmatrix} \begin{bmatrix} n_{c,A}[A]_c \\ n_{c,A}[B]_c \\ n_{c,A}[C]_c \\ n_{c,A}[D]_c \end{bmatrix}$$

What is also important to note is that since the designed graph contains the actual reaction rates for the considered reaction scheme, it represents a slice of the carbon flow pathway at a particular location within the flame. The matrix showing the relationship between the produced and consumed species is called the adjacency matrix.

- **Graph routing**

  The routes on a graph can be characterized as a "walk" or a "path", where a "walk" represents a sequence of edges, whereas a "path" is a walk between two specific nodes, such that the nodes in between are only visited once. Paths, and especially acyclic paths (starting and ending at different nodes), are particularly important because various algorithms for their quantification in complex graphs have been developed. The length of a given path is either the number of edges within the path (for an unweighted graph) or the summation of its weights. The shortest path is the path ensuring a minimum walk.

  For the weighted directed graph shown before, the shortest path would calculate the smallest weighted flux through the path. Since there is no physical interpretation for this shortest distance, we rescale our graph weights so that the shortest distance from A to B corresponds to the maximum flux through all paths into B. To achieve that, we first normalize our edge weights so that the total carbon flux into any given species is 1. This implies that if the original carbon flux for edges were given by $A_{c,i\to j}$ for a given pair of nodes $i, j$, the normalized fluxes would be:

$$K_{e,i\to j} = \left( \frac{A_{e,i\to j}}{\sum_{k,k\neq j} A_{e,k\to j}} \right)$$

  Interestingly, while the addition of these normalized fluxes through a path does not mean anything significant, their product through a path denotes the fraction of carbon starting from node *i* that would reach node *j* while passing through this path. This is a meaningful quantity that finds the dominant pathways through which carbon is being shuffled in the chemical reaction network. To achieve the maximum of this fraction using available shortest path formulations, a new graph is constructed with the same structure as the original, but with edge weights given by:

$$F_{e,i\to j} = -\log\left(\frac{A_{e,i\to j}}{\sum_{k,k\neq j} A_{e,k\to j}}\right)$$

The above formulation allows the use of the established Dijkstra's algorithm (Dijkstra, 2022) since it rescales the edge weights to non-negative values. A visual representation of the above graph is shown below. Interestingly, for some of the edges, the edge weights are zero, indicating that all the flux into that node comes from that edge.

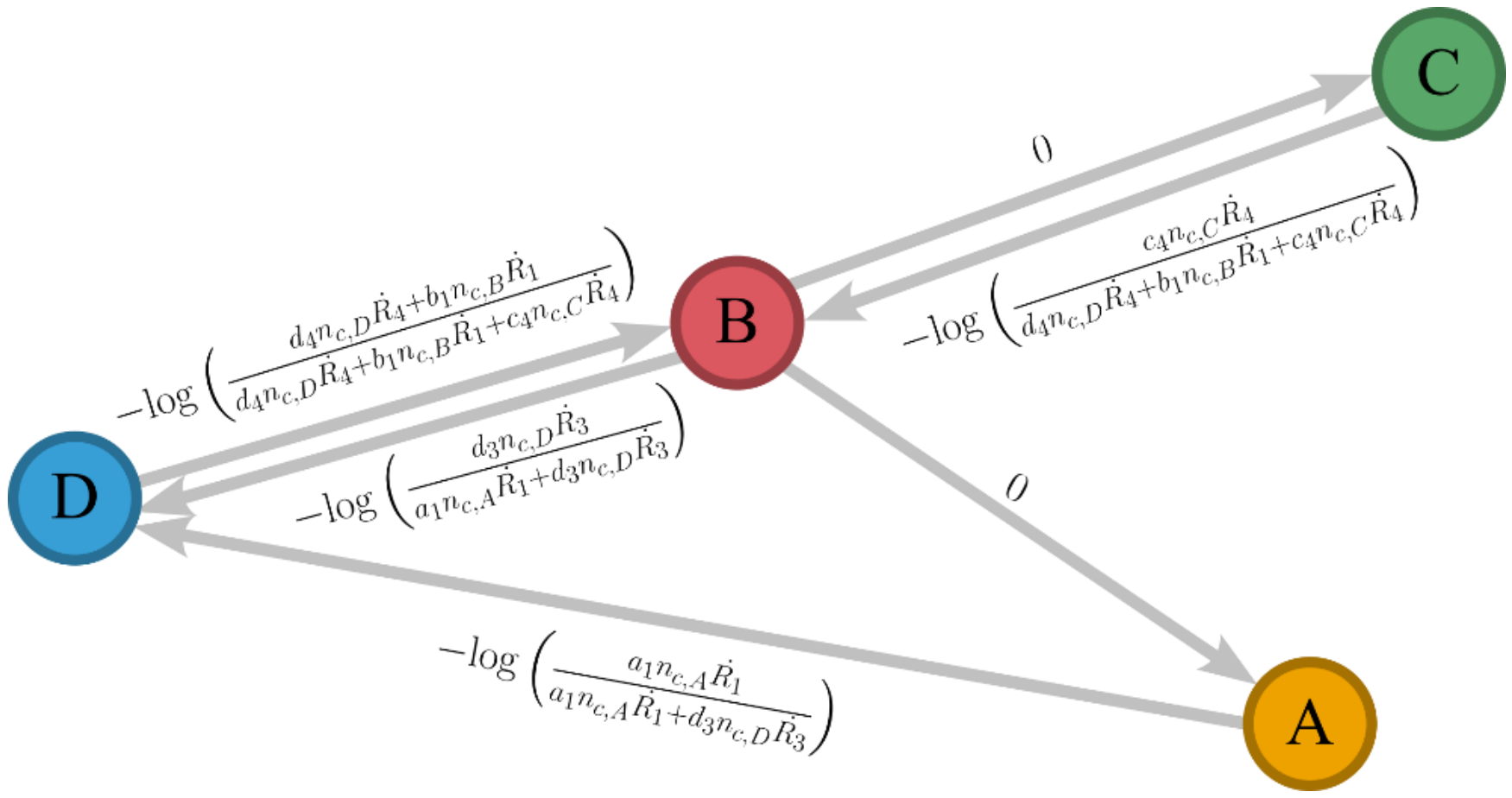


*Figure S5: Transformed weighted digraph which can be used to find the normalized carbon flux through a path using distance metrics.*

Let $p_{ij,k} = (i, \dots, j)_k$ denote the $k^{\text{th}}$ ordered path connecting species $i$ to species $j$. Under this renormalization, the path distance is given by

$$d_{i\to j, p_{ij,k}} = \sum_{l=1}^{N_p - 1} F_{\mathrm{C}, p_{ij,k}(l) \to p_{ij,k}(l+1)}$$

This distance represents the fractional carbon flux delivered to $j$ along $p_{ij,k}$, where $N_p$ is the path length. Essentially, if the total carbon flux into species $j$ is unity, the flux through the path is given by $f_{i\to j, p_{ij,k}} = \exp\left(-d_{i\to j, p_{ij,k}}\right)$.
The shortest path then involves finding the 1st path in the set of ordered paths between $i$ and $j$ path, $p_{ij,1}$ , starting from node $i$ and ending at node $j$, which minimizes the following function and consequently, maximizes the normalized soot flux along its path, $f_{i\to j, p_{ij,1}}$. Following this, we can also define the geometric mean of the normalized flux to a given node, or throughput, $\mathcal{T}_j$, as:

$$\mathcal{T}_j = \left( \prod_{i=1}^{n, i \neq j} f_{i \to j, p_{ij,1}} \right)^{1/n}$$

- **Degree distribution**
  The degree of a node is the number of connections (edges) that the node has to other nodes. For a directed graph, the degree can be classified as an in-degree and an out-degree, demarcating the number of edges pointing towards and away from the node, respectively. The degree distribution of a graph gives the probability of a node having a degree k. The most typical way of presenting this information is using a cumulative distribution function, which plots the probability that a node has a degree smaller than k. In terms of network statistics, it conveys the global topological information. The degree distribution is not only important in the study of real-world networks, but also in their modeling.

  Consider the simplest network model of a random graph (Erdős–Rényi model) where any given edge is included in the graph with a probability p, independently of all other edges. This generates a graph with a binomial degree distribution. Interestingly, these random graphs were noted to have certain important properties about network connectivity and percolation transitions depending on the size of the network and the probability p, as studied by (Erdős & Rényi, n.d.) in 1960. While it is easy to model and serve as an important template for comparing other complex graphs, most networks in the real-world have significantly different degree distributions when compared to the Erdős–Rényi model. The connection between real-world entities typically follows certain physical, chemical and social laws. Modeling these laws and finding their projections in the graph theory is a foundational concept in modern complex system analysis.

  Typical real-world networks are skewed so that a large number of nodes have a few edges, while a few nodes have a significant number of connections (edges). These nodes are typically hub nodes that connect multiple different regions of the graphs. Some networks, like the World Wide Web, have a power law degree distribution, which implies that the probability of a node having degree k is given by $p(k) \propto k^{-\gamma}$. These networks are called scale-free because higher moments, typically defining the spread of degrees around the mean (standard deviation and skewness), diverge for most observed power law degree distributions. This implies that the degree of a node chosen at random cannot be estimated. Scale-free networks exhibit a small-world phenomenon where seemingly unconnected parts of a network can be reached with a relatively small degree of separation, aided by the hub nodes. Graph

models inoculating 'preferential attachment', i.e., connected nodes tend to make more connections, can be used to generate the scale-free topology. These networks are also extremely tolerant against failures where random nodes are disconnected (attacked) and can exhibit a phase transition where increasing node connectivity beyond a certain threshold leads to the creation of giant connected components (percolation).

While several networks have been identified to display a scale-free property, networks with an exponential degree distribution have also been identified recently (Deng et al., 2011). This characteristic has been noted in non-equilibrium or temporal networks where the growth happens through random connection to existing nodes, but preferential treatment is hampered. These networks have a well-defined scale since all higher moments converge and typically exhibit a small average path length (small-worldness). In a recent work (Mutlay & Restrepo, 2015) studying the reaction networks in pyrolytic and combustion chemistry, an exponential degree distribution was identified. In our analysis, it is also found that the chemical reaction networks for high-pressure counterflow flames follow either an exponential degree distribution or a weakly scale-free distribution. The classification of a network as scale-free has been demarcated in a recent work (Broido & Clauset, 2019) with specific criteria for not-scale-free, super-weak, weakest, weak, strong and strongest scale-free networks. For weighted directed graphs, the procedure involves 1) applying 3 different thresholds to generate 3 unweighted graphs, 2) generating 3 degree distributions from each, considering the in-degree, out-degree and degree of nodes within each unweighted graph, and 3) fitting the generated degree distribution to multiple distributions. Finally, the classification is done as follows (Broido & Clauset, 2019):

- Super-Weak: For at least 50% of graphs, no alternative distribution is favored over the power law.
- Weakest: For at least 50% of graphs, a power-law distribution cannot be rejected ($p \geq 0.1$).
- Weak: Requirements of Weakest, and the power-law region contains at least 50 nodes ($n_{tail} \geq 50$).
- Strong: Requirements of Weak and Super-Weak, and $2 < \hat{\alpha} < 3$ for at least 50% of graphs.
- Strongest: Requirements of Strong for at least 90% of graphs, and requirements of Super-Weak for at least 95% of graphs.
- Not Scale-Free: Networks that are neither Super-Weak nor Weakest.

Based on these classifications, the original flux graph was noted to have the weakest scale-free characteristics at downstream distances of soot evolution. Since the soot subgraph contained $\approx$ 100 nodes at most locations, it could also

only display the weakest scale-free characteristics under the rigorous conditions defined here.

- **Backbone and subgraph extraction**
  With the increasing size of graphs, backbone extraction is a key tool for simplifying the analysis of complex networks and drawing tractable conclusions. Large reaction-kinetics graphs often contain hundreds of species and thousands of edges, and without reduction, the underlying structure is obscured by sheer density. In addition, the construction of these graphs naturally produces edge weights that span several orders of magnitude. These extreme weights can distort centrality measures, clustering behaviour, and pathway identification, even though they do not necessarily correspond to mechanistically important carbon-transfer routes. Moreover, since our goal is focused on the soot formation within these network structures, the unreduced graph can be both visually and analytically misleading.

  Backbone and subgraph extraction necessarily removes the unnecessary edges and nodes and, in our case, retains the subgraph, which is important for the carbon transfer to soot. While subgraph extraction can be achieved using various techniques involving the filtering of nodes or edges based on significant or thresholds, their relevance for the study of a particular node (hereby soot) within the context of distance might remain unimpressive. A review of various backbone extraction techniques and their advantages and disadvantages was recently studied (Yassin et al., 2023).

  The backbone extraction in our case is principally similar to the Metric distance backbone extraction technique (Simas et al., 2021), except that instead of selecting the union of all shortest paths, we do a distance-based thresholding of our graph. Essentially, all pathways that contribute at least $\tau = 0.001\%$ of the total carbon flow to soot are selected, while the rest of the edges and their nodes are removed. Note that at any given location within the flame, the threshold does not constitute an absolute flux threshold below which the pathways are ignored. Rather, it is a threshold of the relative flux flowing to soot at that location. This ensures that even though the carbon flow to soot is lower, we select a proportionally lower threshold for the soot subgraph extraction. The sensitivity to this value is analyzed in a different section of this supplementary document. It was noted that for a sufficiently small value of the relative threshold, $\tau \leq 1e-5$, the analyzed centrality measures and soot metrics did not change appreciably. Here, we show the complete graph in cyan and the extracted soot subgraph in red at one of the representative locations of our A1 case flame.

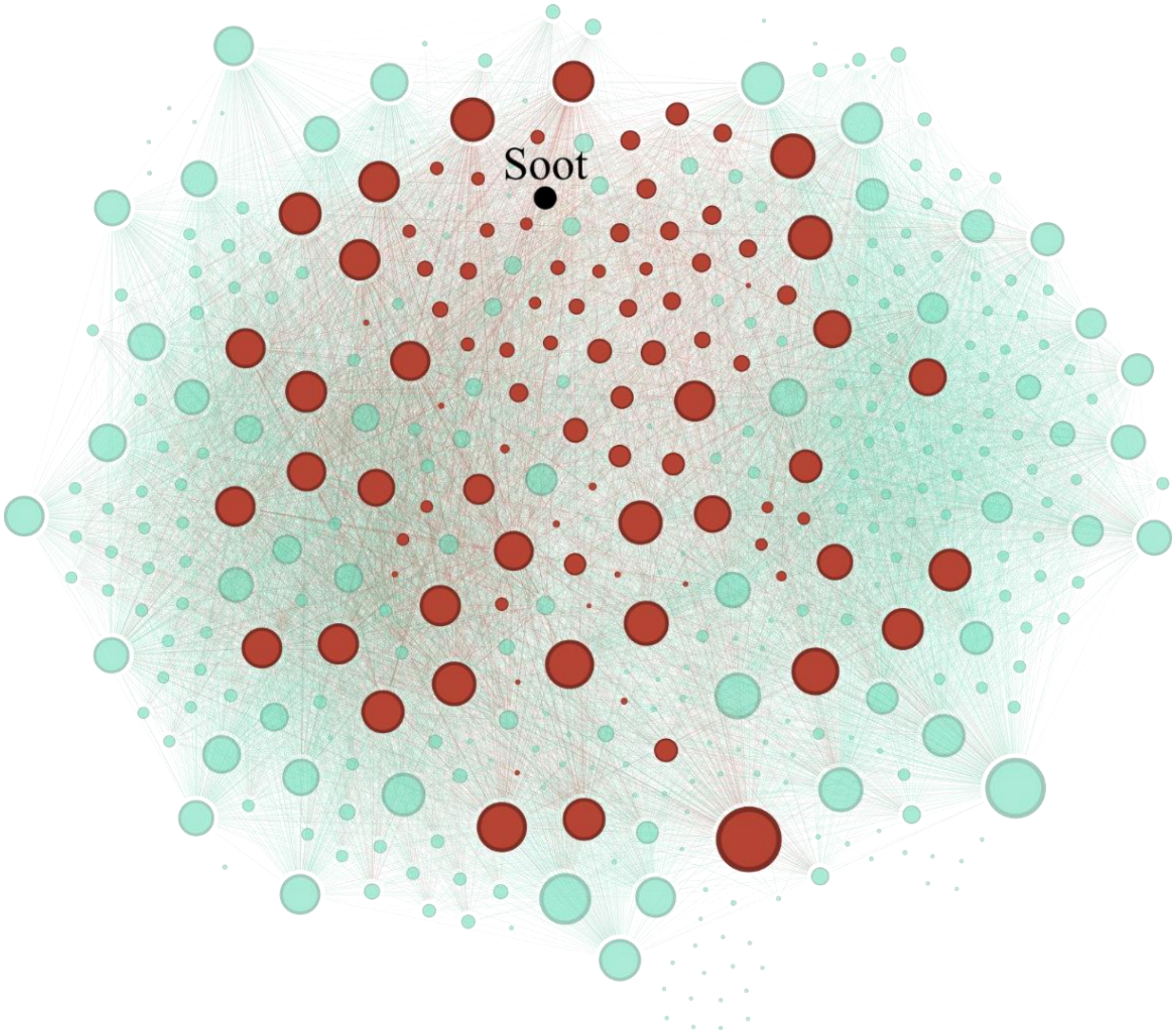


*Figure S6: Complete generated graph at one of the representative locations within the A1 flame (cyan color nodes and edges), along with the extracted soot graph (maroon nodes and edges)*

- **Centrality**
  Within the context of network and graph theory, centrality answers the question "Which node is important?" by essentially ranking the nodes using a defined importance measure. Naturally, the importance of a node can be assessed using various centrality measures. Degree centrality is one of the easiest ones to understand since it ranks the nodes based on their number of degrees. The node with the highest degree ranks highest, while nodes with no connections rank at the bottom. Within the context of chemical reaction networks, centrality measures can determine the hub species within a given context. Since we are interested in soot, we use two specific measures in this work.

  1. **PageRank**
     The classical PageRank (PR) measure was developed by Google to rank the web pages using their link structure. Over time, however, PR has found

importance in the analysis of various graphs and network data. Mathematically, PR can be evaluated as follows.

$$\mathrm{PR}(i) = (1-\alpha)/N + \alpha \sum_{j \in B_i} \left( \mathrm{PR}(j) a_{ji} / L_w(j) \right)$$

Where $A = (a_{i,j})$ is the adjacency matrix of the directed weighted graph, $N$ is the total number of nodes, $B_i$ is the set of all nodes that have an edge into the node $i$, $L_w(j)$ is the sum of the weights of outgoing edges from node $j$, and finally, $\boldsymbol{\alpha}$ is a damping factor representing the probability that a carbon atom continues to follow edges within the graph and does not magically teleport to a node. The damping factor was originally used to model a surfer who can go to a random link without necessarily following the links from his web walk. In the case of carbon transfer to soot, however, the teleportation is unnecessary and has been omitted so that $\alpha = 1$. The updated PageRank definition then makes it so that the surfer (carbon atom in our case) follows a Markov chain process. The updated PR then gives the stationary probability distribution of a random walk. Essentially, if we let carbon flow through our graph network for a sufficiently long time, the probability of finding a certain carbon atom at a given node would be given by $\mathrm{PR}(i)$

2. **Influence**
   Influence is a centrality measure defined in this work for quantifying the importance of a species in carbon flux to soot. In a previous work by the authors, we evaluated the dominant soot formation pathway, which gave the largest carbon flow to soot while starting from the fuel species. Naturally, most of the carbon flow to soot would not flow through the fuel species at a downstream location within the flame. The influence metric was devised to improve this analysis and find the dominant backbone graph through which soot gets most of its mass. The methodology assigns the influence of a node, $I_i$ , as the fraction of carbon that reaches soot while flowing through the given node, $i$.

**SM2: Model validations**

Since we have used the numerical model for the graph-based analysis, it is important to compare its results with previous and current experimental measurements. This would help in ascertaining the model capabilities and limitations of the graph-based analysis. Please note that some of these validation results have already been presented in the supplementary material of another work by the authors (Sawanni & Gülder, 2026). Here, we present the comparison of temperature, soot concentration, and species profiles for

various major and minor species (including PAHs up to pyrene) in high-pressure CDFs. The validation of soot concentration allows us to understand the capabilities of the soot model and is compared using the measured soot volume fractions in this work and from the results of Amin et al. (Hafiz Muhammad Fahid Amin, 2018). In the absence of species measurements under our experimental conditions, we use the results from (Gleason et al., 2021) for comparison.

- **Soot concentration and production rates (current data)**
  The soot concentration and production rates are shown for all the test conditions explored in this work. It can be noted that the peak soot concentration is captured well by the numerical model for the complete range of peak flame temperatures and pressures explored. Deviations may be caused by finite spatial resolution and in-scattering contributions, as described in previous works (Sawanni & Gülder, 2025).

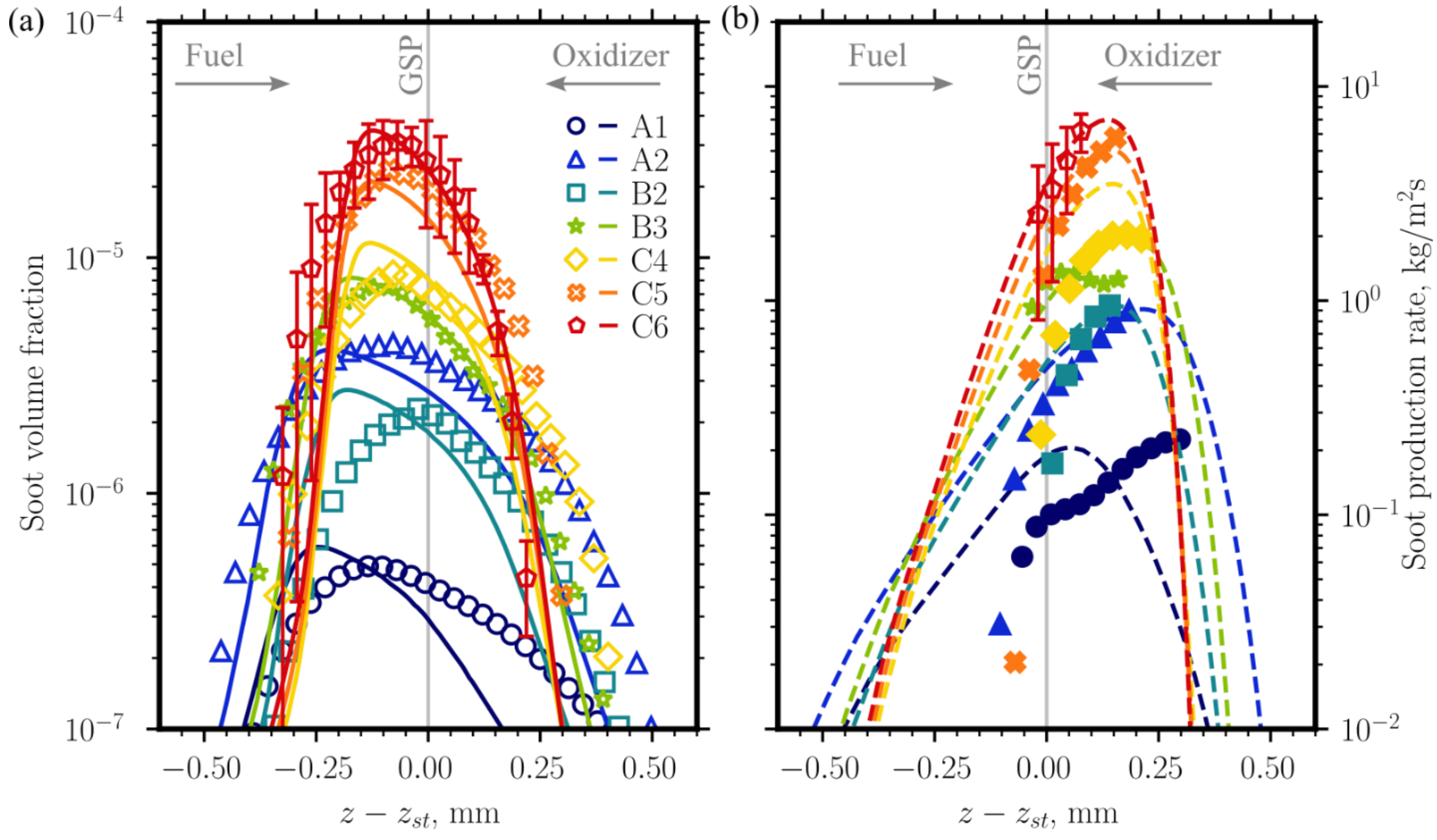


*Figure S7: Experimental and numerical profiles of (a) Soot volume fraction and (b) Soot production rates in the studied cases of this work. Lines represent the computational solution, and symbols show experimental measurements.*

- **Soot concentrations (Hafiz Muhammad Fahid Amin, 2018)**
  While soot concentrations were well replicated in the currently studied experiments, we also compare the model results for previously published data from Amin and Roberts (Amin & Roberts, 2017). The comparison of numerically evaluated and experimentally measured soot concentrations is shown in Figure S8. The model is noted to overpredict the soot concentrations by a factor of 4 at incipient conditions. This might be attributed to deviations in the soot absorption function,

which was not accounted for by Amin and Roberts (Amin & Roberts, 2017). At higher pressures, however, the scaling of soot concentrations is well reproduced. The spatial profile of SVF is similar to that of the current work and typical of a soot-forming CDF. Low soot concentrations near the high-temperature zone are measured for high-pressure flames, but it is likely an artifact of the inversion process.

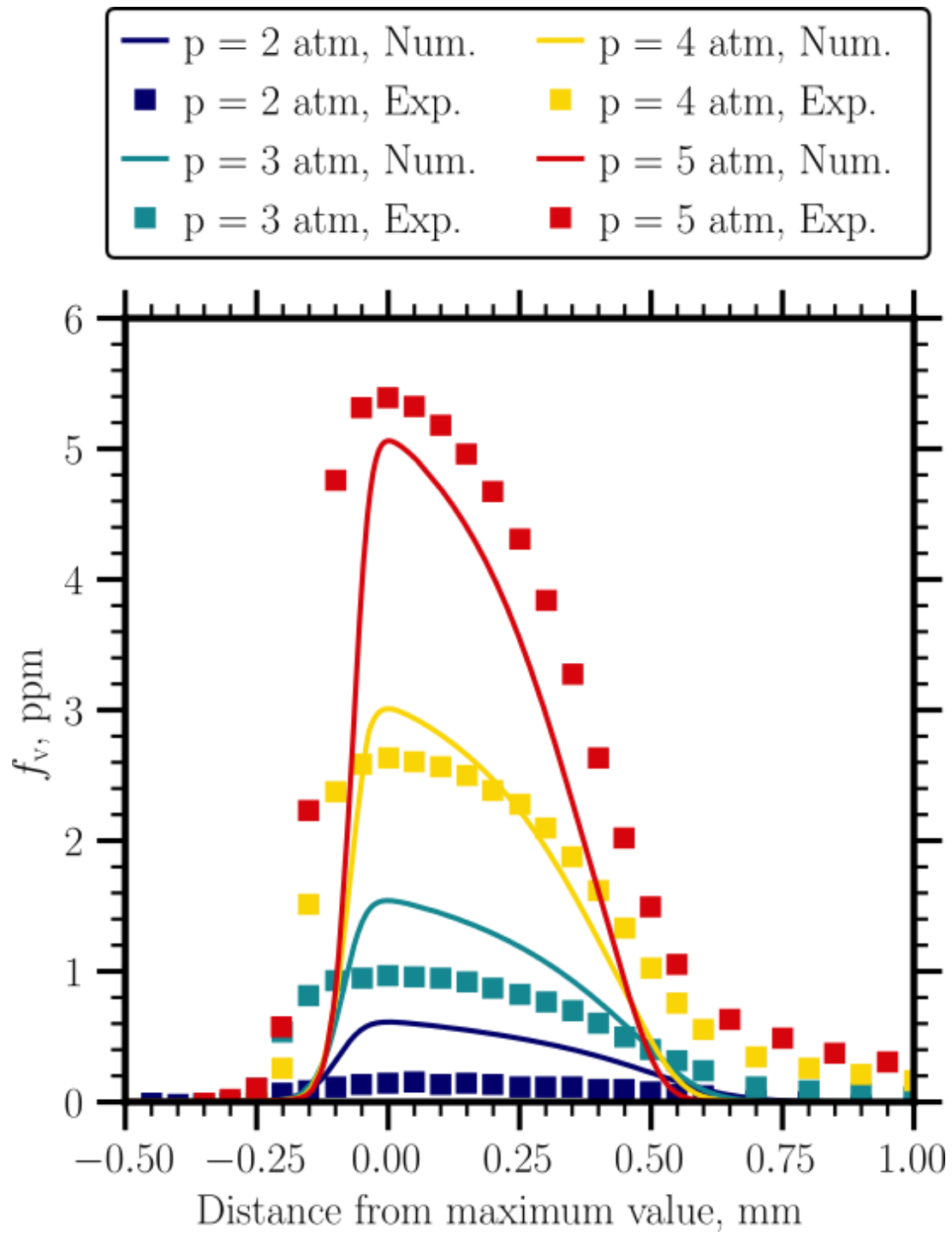


*Figure S8: Measured (Hafiz Muhammad Fahid Amin, 2018) and evaluated soot volume fraction along the axis of counterflow diffusion flame at high pressures*

- **Species concentrations**

Although good agreement with experimental data is achieved for soot volume fraction and temperature, this does not ensure accurate prediction of intermediate processes. Therefore, validation of PAH species remains essential. This data has been presented in a previous work by the authors (Sawanni & Gülder, 2026), but is repeated here for completeness of the numerical validation study. In the absence of direct experimental measurements of PAH concentrations, we rely on the data reported by (Gleason et al., 2021), the boundary conditions for which are listed in Table S1.

*Table S1: Test conditions for validation of temperatures, major and minor species from (Gleason et al., 2021), $X_f$: Fuel mole fraction, $X_{ox}$: Oxygen mass fraction, $U_f$: Centreline local axial velocity at the fuel nozzle exit, $U_{ox}$: Centreline local axial velocity at the oxidizer nozzle exit, $V_{r,f}$: radial velocity at the fuel exit centreline, $V_{r,ox}$: radial*

*velocity at the oxidizer exit centreline, $T_f$: Temperature at the fuel exit, $T_{ox}$: Temperature at the oxidizer exit, p: Pressure, $T_{max}$: Computed peak flame temperature*

| $x_f$ | $x_{ox}$ | $U_f$ m/s | $U_{ox}$ m/s | $\frac{dV_{r,f}}{dr}$ /s | $\frac{dV_{r,ox}}{dr}$ /s | $T_f$ K | $T_{ox}$ K | p atm | $T_{max}$ K |
|---|---|---|---|---|---|---|---|---|---|
| 0.330 | 0.229 | 0.278 | 0.341 | 26.4 | 16.2 | 298 | 323 | 1 | 2013 |
| 0.255 | 0.176 | 0.234 | 0.283 | 17.4 | 9.15 | 298 | 323 | 4 | 1826 |
| 0.225 | 0.155 | 0.216 | 0.259 | 16.2 | 8.54 | 298 | 323 | 8 | 1696 |

Figure S9 compares experimental and numerical profiles of key species and temperature. The left panel presents major reactants, products, intermediate pyrolysis species, and temperature, all of which are well captured by the numerical model. The right panel highlights soot precursors, namely acetylene and PAHs from benzene to pyrene. The acetylene profile, including its scaling with higher pressure, shows excellent agreement with experimental measurements. The profiles of benzene ($C_6H_6$) and naphthalene ($C_{10}H_8$) are in good agreement, though slightly over-predicted by the model. For larger PAHs such as phenanthrene and pyrene, concentrations are available only at atmospheric pressure and are over-predicted by the numerical model, by approximately a factor of 3 for phenanthrene and 10 for pyrene near the stagnation plane.

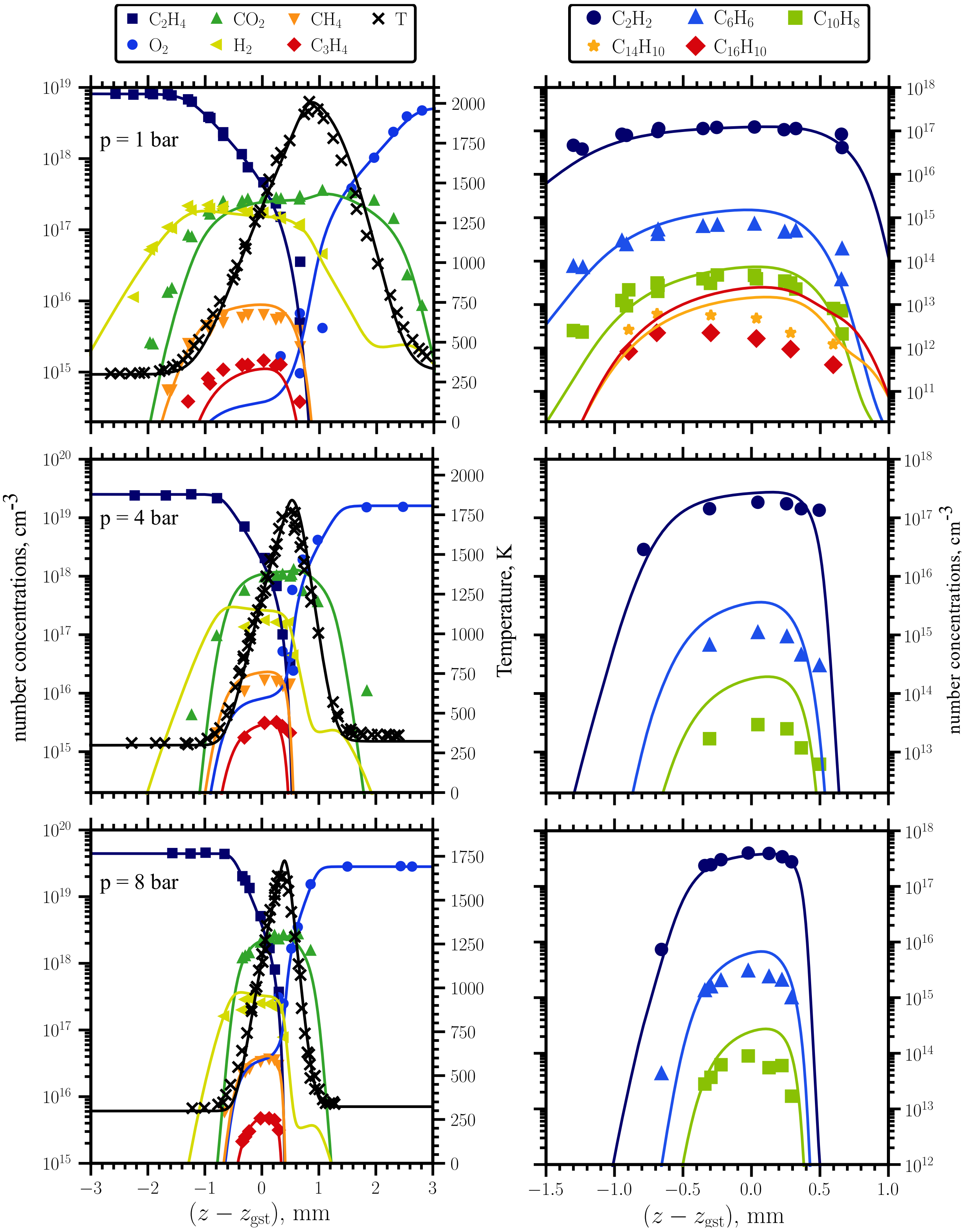


*Figure S9: Comparison of experimental and numerical profiles for Left: reactants, products, temperature, and smaller pyrolysis products, Right: Polycyclic Aromatic Hydrocarbons (PAHs) and acetylene across 1, 4, and 8 atm.*

## SM3: Error analysis

- **Soot volume fraction**

Using the Rayleigh– Debye–Gans/polydisperse-fractal aggregate (RDG/PFA) theory of particle light scattering and extinction for primary particles in the Rayleigh range, that is, $\pi\frac{d}{\lambda} < 0.3$, soot volume fraction is calculated from local extinction data as:

$$f_v = \frac{k_\lambda \lambda}{6\pi E(m_\lambda)(1+\alpha_\lambda)}$$

where $k_\lambda$ is the local extinction coefficient at wavelength $\lambda$, $E(m_\lambda)$ is the soot absorption function, and $\rho_{sa}$ is the scattering to absorption ratio. The uncertainty in the SVF can then be estimated based on the error propagation equation as:

$$\sigma_{f_v}{}^2 = \left(\sigma_{k_\lambda}\frac{\partial f_v}{\partial k_\lambda}\right)^2 + \left(\sigma_\lambda \frac{\partial f_v}{\partial \lambda}\right)^2 + \left(\sigma_{E(m_\lambda)}\frac{\partial f_v}{\partial E(m_\lambda)}\right)^2 + \left(\sigma_{\rho_{sa}}\frac{\partial f_v}{\partial \rho_{sa}}\right)^2$$

$$\left(\frac{\sigma_{f_v}}{f_v}\right)^2 = \left(\frac{\sigma_{k_\lambda}}{k_\lambda}\right)^2 + \left(\frac{\sigma_\lambda}{\lambda}\right)^2 + \left(\frac{\sigma_{E(m_\lambda)}}{E(m_\lambda)}\right)^2 + \left(\frac{\sigma_{\alpha_\lambda}}{(1+\alpha_\lambda)}\right)^2$$

where $\sigma_{f_v}, \sigma_{k_\lambda}, \sigma_\lambda, \sigma_{E(m_\lambda)}, \sigma_{\alpha_\lambda}$ are the standard deviation uncertainty in SVF, local extinction coefficient, wavelength, soot absorption function, and scattering to absorption ratio, respectively. Since $k_\lambda$ is the only measured quantity and the source of precision uncertainty, the uncertainties in $f_v$ due to $\sigma_\lambda, \sigma_{E(m_\lambda)}$ and $\sigma_{\rho_{sa}}$ scale with $k_\lambda$, and are marked as relative uncertainties. It is noted here that while the soot absorption function is also being evaluated experimentally using the measurement of dispersion exponents and relations of Bescond et al. (Bescond et al., 2016), the precision uncertainty in its values remains dependent on the measured extinction and emission coefficients at multiple wavelengths. The uncertainty in the local extinction coefficient is a function of the actual uncertainty in the measured transmission values, which is further amplified in the centerline by the Abel inversion process (Daun et al., n.d.). The uncertainty in the line-of-sight extinction coefficient is given by:

$$\sigma_{\ln(1/\tau)} = \sqrt{\frac{1}{n}\left(\left(\frac{\sigma_{I_0}}{I_0}\right)^2 + \left(\frac{\sigma_I}{I}\right)^2\right)}$$

Where $\sigma_{\ln\left(\frac{1}{\tau}\right)}$ is the uncertainty in the line-integrated extinction coefficient, $\sigma_{I_0}$ is the uncertainty in background light intensity, $\sigma_I$ is the uncertainty in transmitted light intensity, and $n$ is the total number of images acquired for measurement. Finally, as suggested by Dasch (Dasch, 1992), the propagation of precision uncertainty through the inversion process can be written through the relationship:

$$\sigma_{k_\lambda}(r_i) = \frac{\sigma_{\ln(1/\tau)}}{\Delta y}\left(\sum_{j=0}^{n} D_{ij}^2\right)^{1/2}$$

In the current measurements, the relative errors in the background intensity, $\left(\frac{\sigma_{I_0}}{I_0}\right)$, and transmitted intensity, $\left(\frac{\sigma_I}{I}\right)$, was measured to be 1.74% and 1.58% respectively, which was noted to remain roughly uniform across the sensor. The standard deviation in the line-of-sight extinction coefficient can then be evaluated to be $\sigma_{\ln(1/\tau),1} = 0.0018$ for an acquisition consisting of 500 images. Moreover, due to non-uniformity in illumination optics and reflections of the emission field, an additional error in the extinction coefficient can be evaluated. These uncertainties can be gauged from the light extinction in the far edge of the sensor, away from the flame and light-attenuating field. At these locations, an ideal optical system would have no light attenuation. Based on the deviation from the expected zero line-integrated extinction, an uncertainty in the measurement system can be assigned $\sigma_{\ln(1/\tau),2} = 0.005$. The error from line-of-sight projected data can now be combined using a quadrature relation $\sigma_{\ln(1/\tau)} = 0.0053$. This error is known to amplify due to the application of the inverse Abel algorithm, and an error amplification factor of 40 was evaluated for the regularization parameter used in this study. Consequently, a maximum standard deviation of $\sigma_{k_\lambda,p} = 0.21$ was evaluated, which corresponds to a precision uncertainty of 59 ppb for a 95% confidence interval.

The total uncertainty in the measurements is a combination of systematic and precision uncertainties. Systematic uncertainties arise from errors in the estimation of wavelength, absorption coefficients, soot absorption function, and scattering-to-absorption ratios. In addition to precision-related errors, several other systematic factors may affect the measurement of radial absorption coefficients. The challenges associated with line-integrated and path-integrated measurements have been discussed previously, along with the limitations imposed by finite spatial resolution. Beam deflection or steering is another important source of uncertainty, as it can affect line-integrated extinction measurements and, consequently, the inferred absorption coefficients. The present work places particular emphasis on minimizing beam-steering effects through the use of carefully designed optics and a diffuse light extinction diagnostic. The systematic uncertainty ($\sigma_{\mathrm{kext}}$,λ,s) in the extinction coefficient is evaluated using a quadrature relationship:

$$\sigma^2{}_{k_{\mathrm{ext},\lambda},s} = \sigma^2{}_{k_{\mathrm{ext},\lambda},1} + \sigma^2{}_{k_{\mathrm{ext},\lambda},2} + \sigma^2{}_{k_{\mathrm{ext},\lambda},3}$$

Where $\sigma_{k_{\mathrm{ext},\lambda},1}$ is the error associated with line integration errors, $\sigma_{k_{\mathrm{ext},\lambda},2}$ is the error associated with finite spatial resolution and $\sigma_{k_{\mathrm{ext},\lambda},3}$ is the error associated with anisotropic scattering (in-scattering and out-scattering). These uncertainties were

evaluated in a recent work (Sawanni & Gülder, 2025) and are shown for one of our studied cases (C6) in Figure S10. The figure shows the actual and measured or corrected values of extinction and radiance profiles. Because the deviations vary significantly with spatial location, these profiles are used to estimate the local uncertainty in soot concentrations based on the measured extinction coefficients.

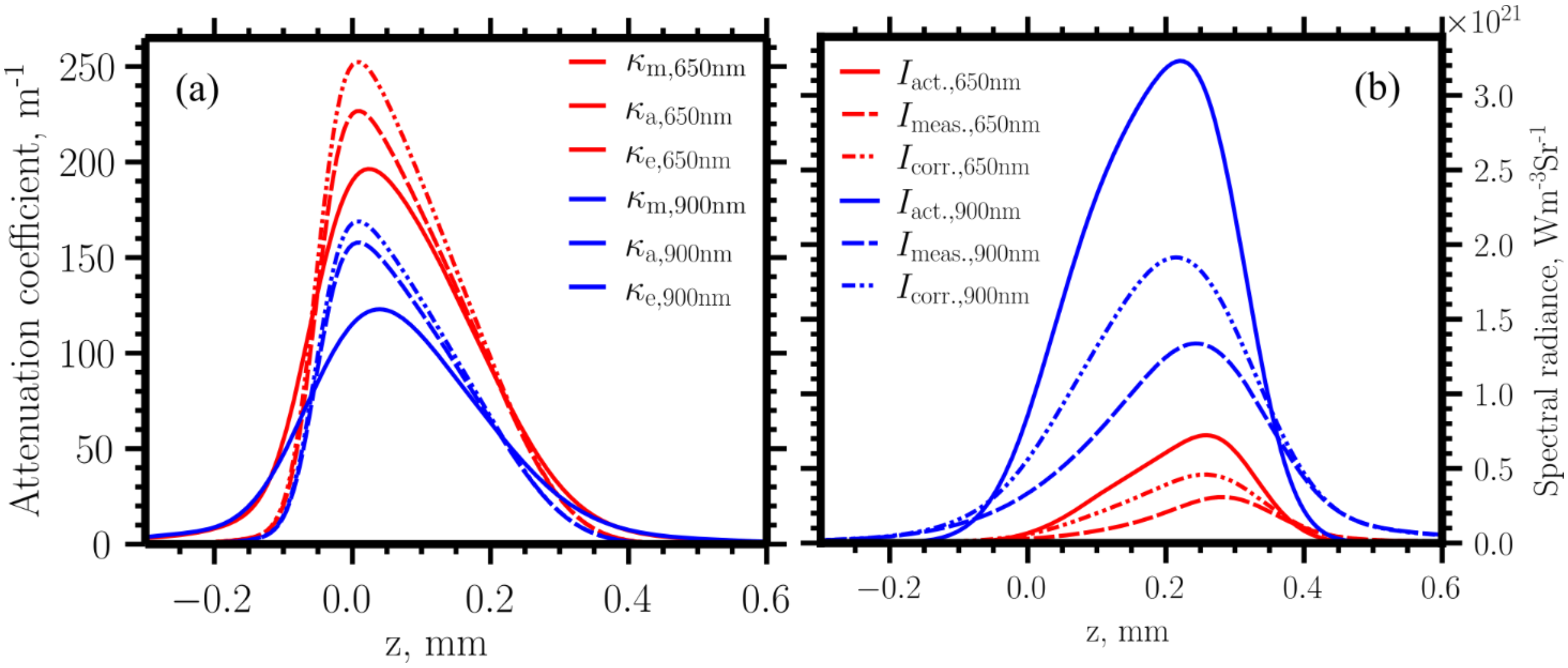


*Figure S10: Influence of finite spatial resolution, anisotropic scattering, and line integration errors on the (a) measured extinction coefficients at 650 nm and 900 nm, when compared to the original extinction and absorption coefficients; (b) measured and self-absorption corrected radiance profiles when compared to the actual radiance emitted by the flame. Both curves are shown for the C6 flame with the following notations $\kappa_{m,\lambda}$, $\kappa_{a,\lambda}$, $\kappa_{e,\lambda}$ are the measured, absorption and extinction coefficients at a given wavelength $\lambda$, while $I_{act,\lambda}$, $I_{meas.,\lambda}$, $I_{corr.,\lambda}$ are the actual, measured and self-absorption corrected radiance profiles.*

With the relative and precision uncertainties in extinction coefficients established, we next consider uncertainties associated with additional quantities. Use of the central wavelength of the bandpass filter provides an accurate estimate of the measurement wavelength; therefore, wavelength uncertainty arises primarily from uncertainty in the central wavelength itself. Based on the manufacturer's specifications, this uncertainty is taken as $\sigma_{\lambda} = 2nm$ for a 95% confidence interval.

The uncertainty in the scattering-to-absorption ratio is difficult to quantify in the absence of detailed particle size distribution and optical property measurements. Based on soot morphology measurements reported by Amin et al. (Amin & Roberts, 2017), the uncertainty in the scattering-to-absorption ratio is conservatively assumed to be $\sigma_{\alpha_{\lambda}} = 0.1$. While this ratio varies with spatial location, reported values do not exceed $\alpha = 0.2$, even for the highest-pressure and most strongly sooting flames examined (Hafiz Muhammad Fahid Amin, 2018). Moreover, the scattering albedo profile typically follows the soot concentration profile in SF CDFs, as observed in the experimental work of Amin (Hafiz Muhammad Fahid Amin, 2018) and the model results used in our previous work (Sawanni & Gülder, 2025). Because

the model does not necessarily capture soot morphology with full fidelity, uncertainties in soot scattering albedo are evaluated using both numerically predicted spatial profiles and a constant value of $\sigma_{\alpha_\lambda} = 0.1$. The larger of the two estimates is adopted as the scattering albedo uncertainty.

Finally, the soot absorption function, E(m), remains one of the most debated quantities in the literature, with reported values varying significantly with soot maturity. In this work, both the wavelength dependence and absolute value of E(m) are quantified, as these are required to accurately infer soot volume fractions from absorption coefficients. This is achieved using experimental data and modeling results from Bescond et al. (Bescond et al., 2016), which relate dispersion exponents to independently measured absorption functions. However, the corrected soot absorption function can be applied only at locations in the flame where both emission and extinction signals are sufficiently strong. As a result, absorption-function corrections are not applied uniformly across all spatial locations. To remain conservative, the full uncertainty in the soot absorption function is included in the uncertainty analysis of soot volume fractions. Based on the work of Krishnan et al. (Krishnan et al., 2000), we estimate the uncertainty in the soot absorption function to be $\sigma_{E(m)} = 0.052$.

- **Temperature**

  The evaluation of temperature in the current work uses the relation:

$$T = \left(-\frac{k_b \lambda}{hc} \ln\left(\frac{\lambda^5 J_\lambda}{2hc^2 k_\lambda}\right)\right)^{-1}$$

where the self-absorption corrected absolute radiance ($J_\lambda$) and measured extinction coefficient at multiple wavelengths are used to evaluate the temperature. The error propagation equation for temperature can be written as:

$$\left(\frac{\sigma_T}{T}\right)^2 = \left(\frac{\sigma_{J_\lambda/k_\lambda}}{J_\lambda/k_\lambda}\left(\frac{T k_b \lambda}{hc}\right)\right)^2 + \left(\frac{\sigma_\lambda}{\lambda}\left(1 - \frac{5 T k_b \lambda}{hc}\right)\right)^2 + \left(\frac{\sigma_{\alpha_\lambda}}{(1+\alpha_\lambda)}\left(\frac{T k_b \lambda}{hc}\right)\right)^2$$

The errors associated with uncertainties in the extinction coefficient and corrected absolute radiance were evaluated in a recent study by the authors (Sawanni & Gülder, 2025). It was noted that the finite and disparate spatial resolutions played the largest role in the observed uncertainty in temperatures, which increased with increasing temperatures. The uncertainty in temperature is noted to depend on the ratio of $J_\lambda$ and $\kappa_\lambda$ and since their errors are noted to be coupled, we estimate the errors in their ratios using the values shown in Figure S10. With the errors in $J_\lambda/\kappa_\lambda$ known with spatial location, the errors in temperature can be estimated by combining them with the uncertainties from wavelength and soot scattering

albedos. For our experiments, the temperature uncertainties are plotted in Fig 2(b) of the manuscript and are noted to be $< 50K$ for all reported locations.

- **Dispersion exponents**
  The evaluation of dispersion exponents is completed using a regression fit for the LHS and RHS of the relation:

$$a + \beta \ln\lambda = \ln\left(\frac{B_\lambda(T)}{J_\lambda \lambda}\right)$$

  Where $B_\lambda(T)$ is the blackbody radiation at a given wavelength and temperature. The errors in dispersion exponents then depend on the measured temperature and emission profiles and their wavelength dependence. Consequently, its uncertainty is propagated from the reported errors in temperature and self-absorption corrected radiance measurements. Since the evaluation of dispersion exponents involves a regression analysis where the effects of spatial resolution, scattering and self-absorption correction vary with wavelengths, we utilize a sensitivity-based error estimate instead of a linear error propagation analysis. The upper and lower bounds of temperature are calculated based on the measured temperature and its uncertainties, which are then translated to the bounds of blackbody radiation at a given wavelength. The systematic uncertainty in self-absorption corrected radiance intensity due to finite and disparate spatial resolution, line integration errors and effects of anisotropic scattering is combined with its precision uncertainty (due to calibration errors quantified at 5%) to generate a range of radiance estimates. The dispersion exponents are regression-fitted from all combinations of these datasets (measured radiance and blackbody radiation) to evaluate a range of dispersion exponents. The errors from these are reported in Fig 2(b) of the manuscript and are noted to be $\sigma_\beta \approx 0.15$ in most locations.

- **Soot production rates**
  The errors in soot production rate estimation stem largely from the uncertainty in the experimentally measured soot volume fractions and the methodology employed to relate experimental and numerical results. The soot volume fractions are first converted to a soot mass fraction using the relation $y_s = f_v \rho_s / \rho$ where $\rho_s$ is the soot density, and $\rho$ is the gas density, which depends on temperature. The dependency of gas density is removed since soot production rates are estimated using a flux balance equation in our work:

$$\dot{\omega}_s''' = \frac{d}{dz}\left(\rho Y_s\left(U_p\right)\right) + \frac{2\rho Y_s v}{r}$$

  where $U_p$ is the particle axial velocity and $v/r$ is the radial velocity gradient, which are both taken from numerical simulations. The uncertainty primarily arises from

matching the numerical and experimental profiles, which is necessary for extracting the numerical terms used in the analysis. In this work, the experimental and numerical profiles are aligned using the location of the PSP, defined as the point where the soot volume fraction (SVF) drops to half of its peak value. Based on this definition, and considering the uncertainty associated with the PSP band for particles of different size classes, we estimate a standard deviation of $\sigma_{z_{psp}} = 0.011\ \mathrm{mm}$, which is verified by the CH* chemiluminescence profiles. Because this uncertainty propagates nonlinearly through the evaluation procedure, we determine error bounds on the soot production rates using a sensitivity study. In this approach, errors consistent with the estimated bounds are added to the experimental SVF, and the perturbed SVF fields are then processed through the soot-production-rate methodology together with uncertainties in the position matching. The mismatch between experimental and numerical fields leads to large uncertainties near the PSP and in the low-sooting, high-temperature region. For this reason, measurement data in these zones are excluded, and only locations with a high signal-to-noise ratio are retained. In the reported zones, the uncertainty in production rates remains below the modest value of 30%.

## SM4: Uncorrected and corrected SVF, SPR profiles

Since the evaluation of the soot absorption function is possible for the range of dispersion exponents noted here, we report the E(m) uncorrected and corrected values of SVF, SPR and compare them to the numerical profiles. The maturity-corrected results show a systematic but moderate impact on the reported quantities. Specifically, soot concentrations increase for lower-pressure flames (e.g., A1), bringing the experimental measurements into closer agreement with the numerical results, while small decreases in SVF are observed at higher pressures. As a result, the overall pressure scaling of soot concentration and soot production rate is slightly reduced compared to the uncorrected values. A direct comparison of the uncorrected and maturity-corrected soot volume fractions and soot production rates is provided in Figure S11.

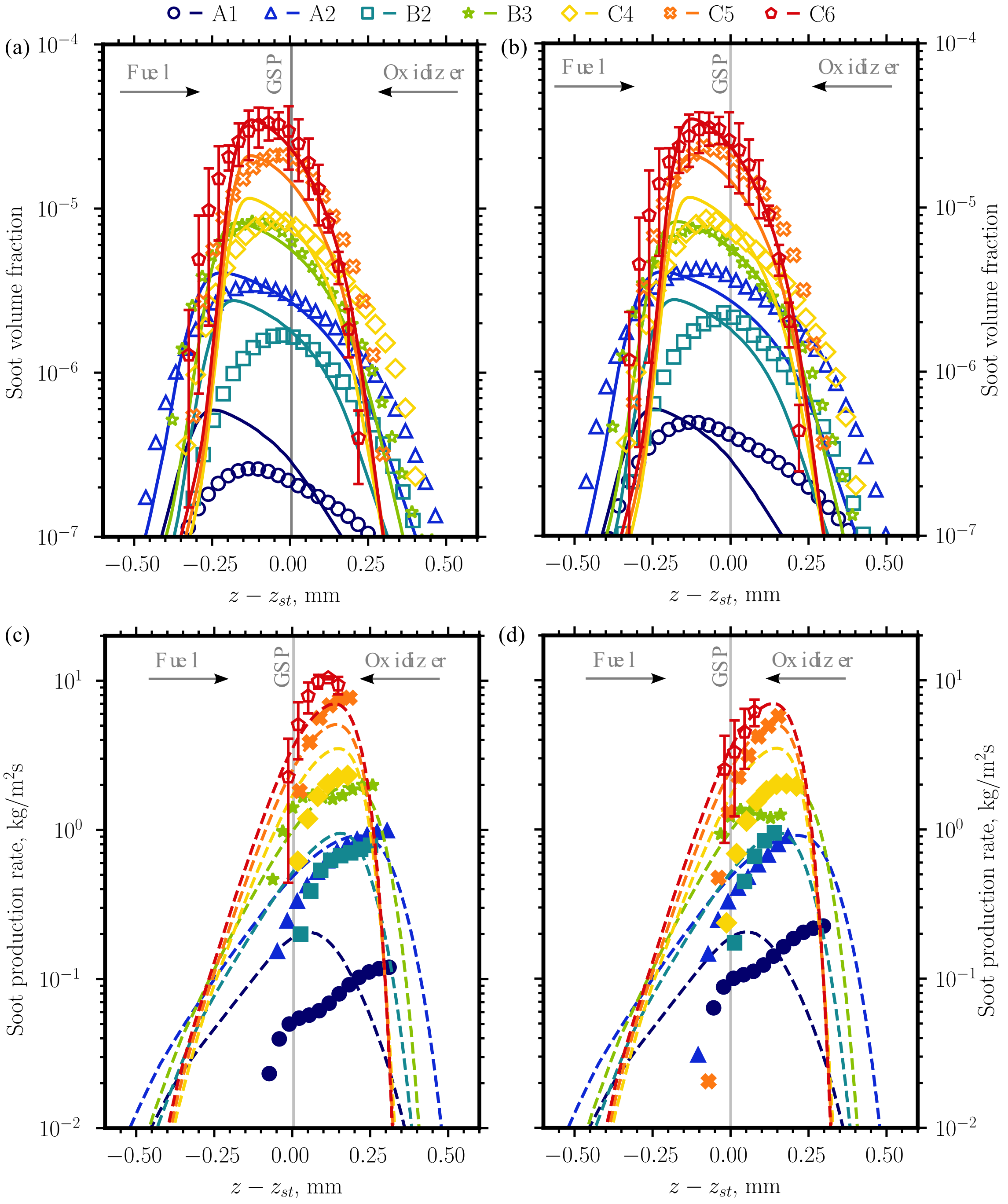


*Figure S11: Influence of maturity-dependent corrections to $E(m_\lambda)$ on soot concentration and soot production rates for the studied flames. (a), (b) Uncorrected and corrected soot volume fractions; (c),(d) Uncorrected and corrected soot production rates.*

## SM5: Sensitivity analysis for graph metrics

The extraction of the soot subgraph necessarily involves thresholding choices. In the present framework, the full reaction network is filtered based on the cumulative carbon flux

reaching soot through each pathway. Specifically, any pathway contributing at least 0.001% of the total carbon flux to soot is retained, while pathways below this threshold are excluded. Importantly, this criterion is defined relative to the total carbon flux to soot rather than on an absolute basis. This ensures that all pathways with a meaningful contribution to soot growth are included in the extracted subgraph.

This thresholding naturally retains different classes of species and pathways. Species closer to soot (e.g., larger PAHs) typically contribute higher individual fluxes but through fewer pathways, while species farther from soot formation (e.g., acetylene) contribute through a much larger number of individually smaller yet collectively significant pathways. The union of all such retained pathways defines the soot subgraph. Here, we perform a systematic sensitivity analysis for all graph-based metrics used in this study. Note that PageRank(soot) and its throughput $\mathcal{T}_s$ is calculated using the original graph without any thresholding. These metrics are consequently insensitive to the choice of thresholds. The soot subgraph and all associated network measures, on the other hand, were recomputed for three different threshold values ($\tau = 0.0005\%, 0.001\%, 0.005\%$) spanning an order of magnitude. Over this range, the sensitivity of the network metrics and the resulting trends was found to be nominal, indicating that the conclusions are not an artifact of the specific threshold choice. For substantially higher thresholds (greater than 1% of total carbon flux), a large number of low-normalized flux but cumulatively important pathways from species farther upstream of soot formation are removed. This disproportionately affects metrics associated with smaller species, such as acetylene.

Since one of the key conclusions of this work is the increasing dominance of acetylene-mediated pathways at elevated pressures, selecting a relatively small threshold is both physically justified and necessary to accurately represent the full carbon flow to soot. Using lower thresholds systematically improves the fidelity of the extracted subgraph in capturing the underlying chemistry, albeit at a significantly increased computational cost due to the exponential growth in the number of enumerated pathways. An appropriate balance between physical completeness and computational tractability is therefore required, particularly for influence metrics that depend on the enumeration of all possible pathways from a given species to soot. The sensitivity of various metrics to the threshold choices is shown below.

- **Degree distribution**
  The degree distribution of the extracted subgraph is shown for three different threshold values in Figure S12. Increasing the threshold removes edges associated with pathways that contribute only modestly to carbon flow toward soot. Consequently, the average degree decreases across all locations. At higher threshold values, the degree distribution of the soot subgraph before the soot inception zone (red shaded lines in the figure) exhibits a sharp exponential decay, with no nodes having a degree greater than 10. As the network evolves, a heavier tail

emerges for all threshold values (blue shaded lines in the figure); however, the degree of hub nodes decreases as the threshold increases. This is likely due to the reduction in pathways involving smaller species, which typically serve as connectors between multiple regions of the graph. Although the quantitative details differ slightly, all networks ultimately develop hub species and display weak power-law characteristics.

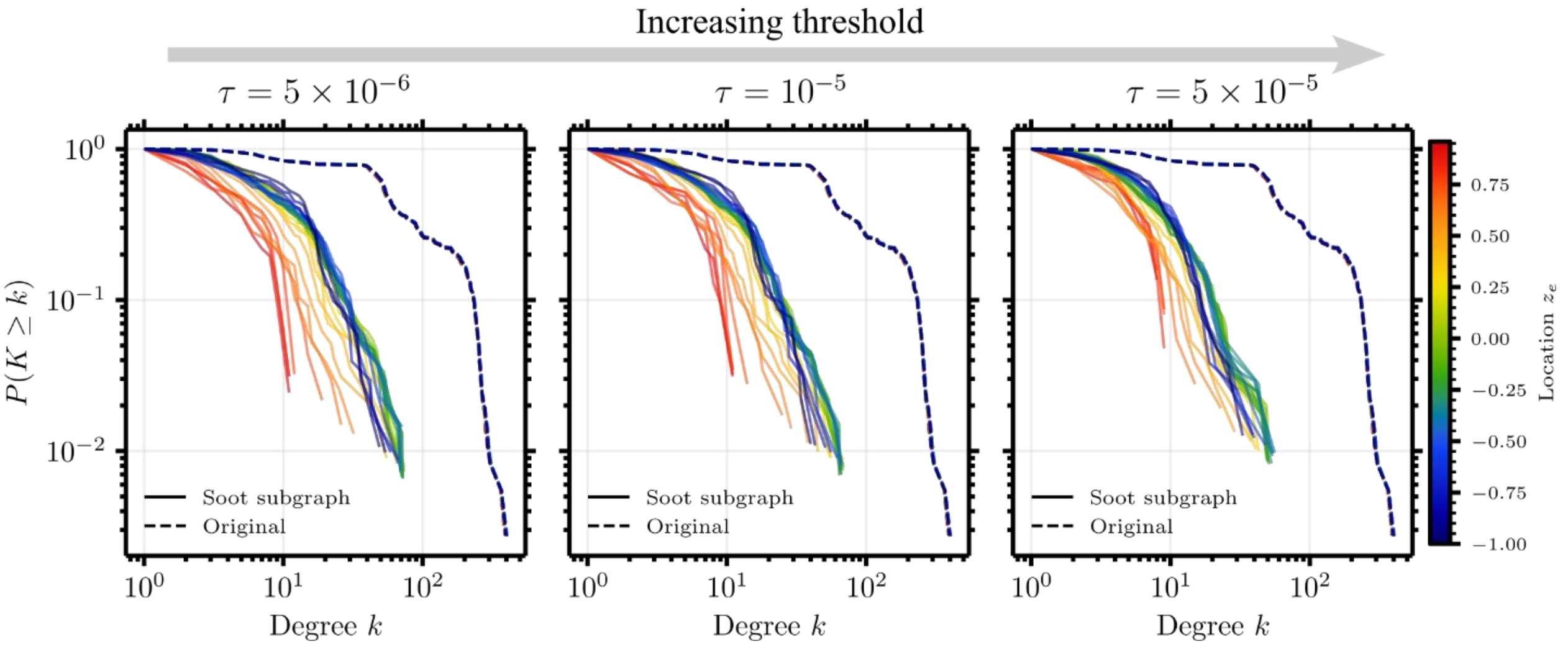


*Figure S12: Effects of threshold values on the degree distribution of the generated soot subgraphs at various locations of the A1 flame. The degree distribution of the original graph is also overlaid.*

- **Graph size and backbone structure**
  The change in graph size and backbone structure is shown in Figure S13. As expected, increasing the threshold value decreases both the number of edges and the number of nodes in the graph. The number of edges decays more rapidly than the number of nodes, since the average degree remains relatively large. Although the graph size and order decrease, the characteristic progression of the network through the soot-forming zone remains consistent. Moreover, the backbone structure, representing the flow of carbon through influential nodes, also remains largely unchanged, particularly in the evolved network near and beyond the growth region on the left side of the spatial domain. The only notable difference appears in the initial soot subgraph, which contains an almost linear chain of species. The branching of this linear chain farther from the soot is sensitive to the choice of threshold. For threshold values lower than the selected value, the conversion of anthracene to the fulvenallenyl radical also appears in the network backbone, while a higher threshold removes the contributions from weak and farther edges. It is noteworthy that addition and subtraction occur along weak links connecting to soot, highlighting their susceptibility to threshold selection. Once the carbon flow to soot strengthens, small variations in the threshold do not affect the carbon-flow structure or the conclusions drawn in this study.

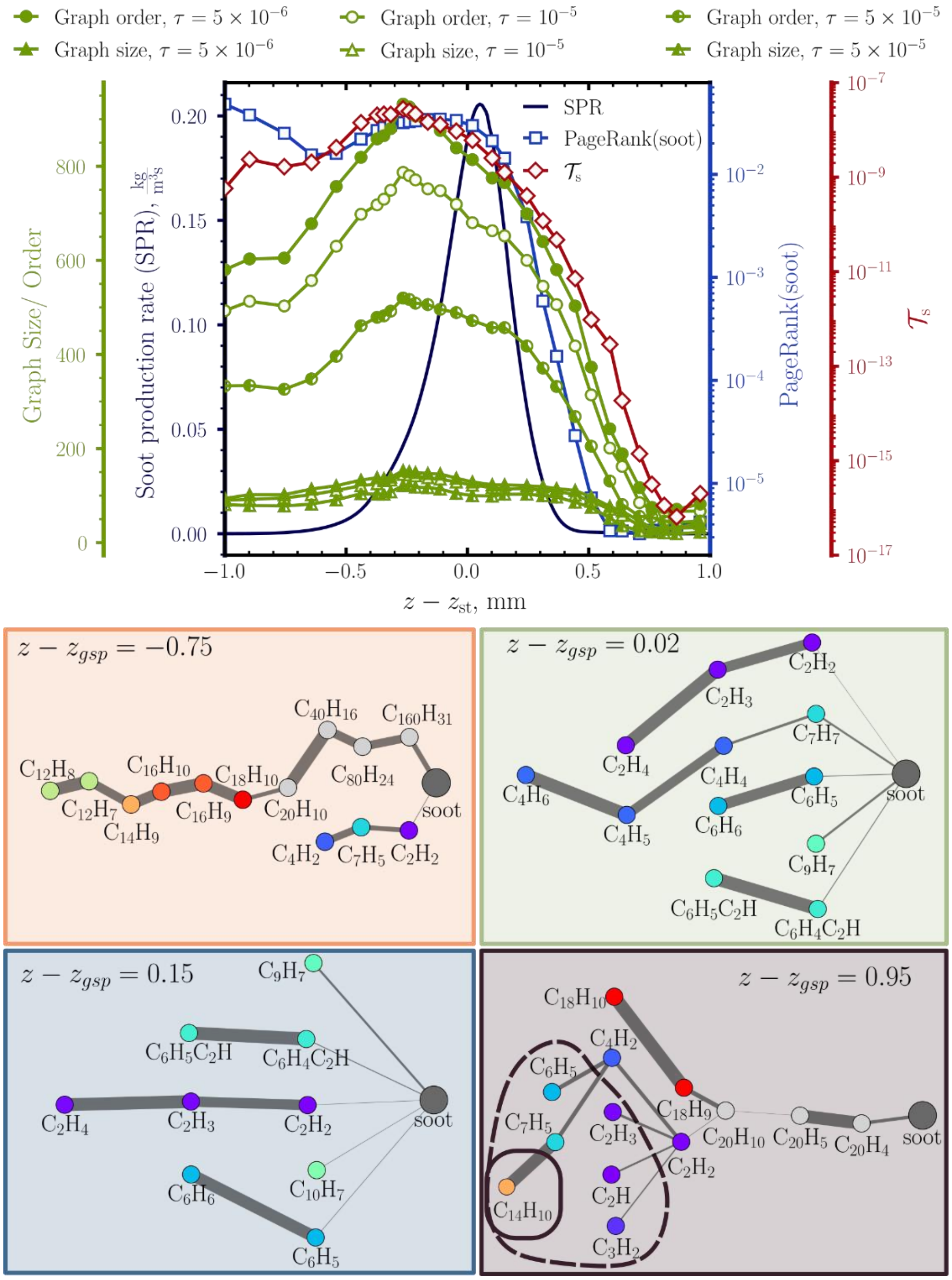


*Figure S13: Effects of threshold on the graph size, order, and backbone structure. In the bottom panel, dashed (addition due to decrease in threshold) and solid (removal due to increasing threshold) boundaries indicate the edges and nodes that are prone to addition or subtraction by changing threshold values.*

The growth of the network and the strengthening of its carbon flow also proceed in a similar manner across different threshold values. Figure S14 shows the probability of attachment of new nodes to existing ones as the soot network evolves, with error bars indicating the range of values obtained under varying threshold choices. While small deviations arise due to the retention or removal of weak edges, the

attachment probability associated with hub nodes remains largely unchanged. This stability suggests that the dominant growth mechanisms of the soot-forming network are robust to moderate variations in threshold selection, at least for the selected range of thresholds. More broadly, the persistence of hub-driven attachment across thresholds reinforces the idea that the network's large-scale structure is governed primarily by strong, chemically meaningful pathways rather than by the numerous weak connections that are pruned at higher thresholds. In other words, although thresholding alters the fine-grained details of the graph, the underlying preferential-attachment-like behavior, where highly connected species continue to attract new connections, remains intact. This consistency further supports the reliability of the inferred growth patterns and the conclusions drawn from them.

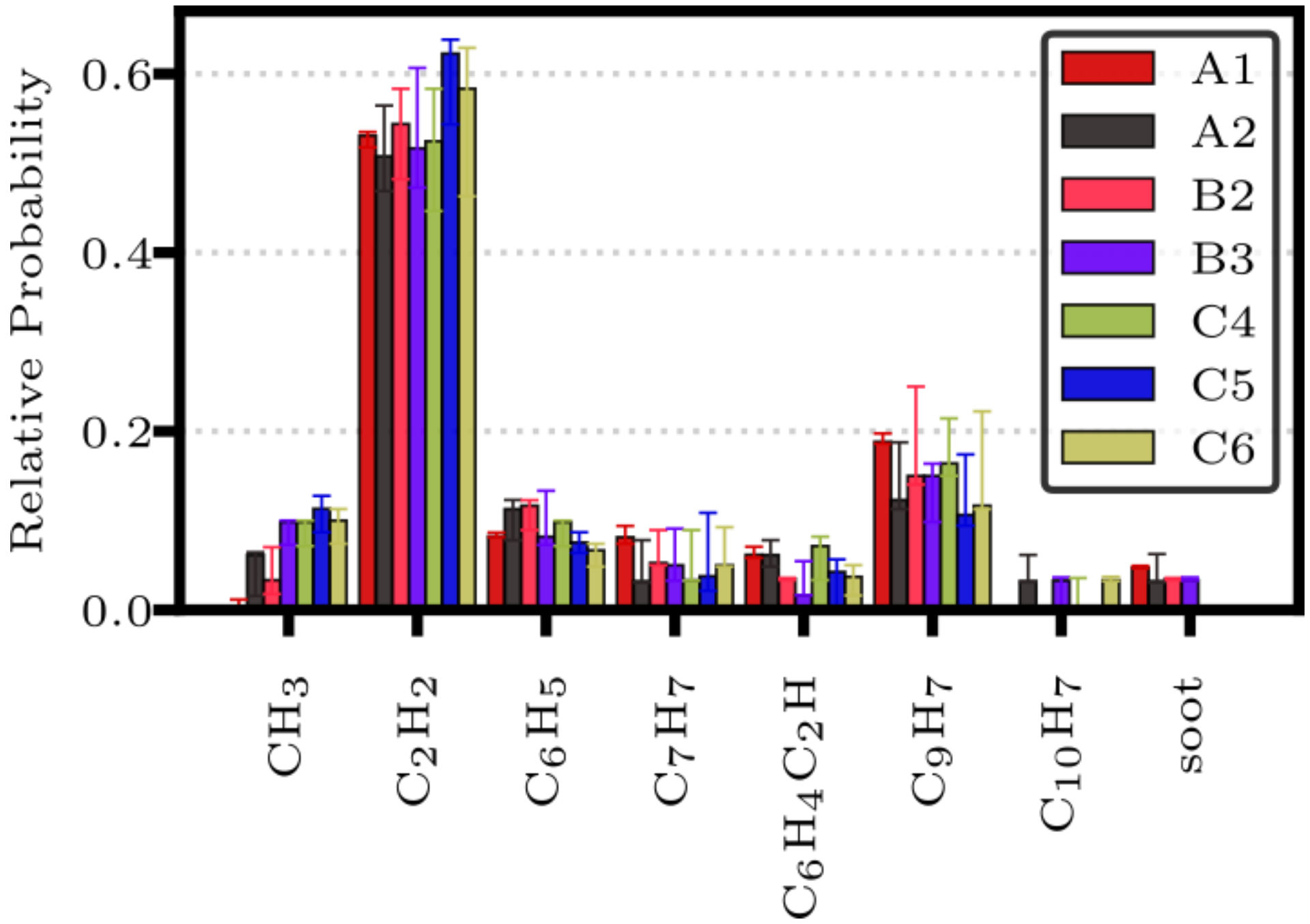


*Figure S14: The probability of attachment of new nodes to the listed hub nodes across the various cases. The analysis is done in the regions of increasing graph size as soot evolves through a temperature-decaying environment. Error bars indicate the range of probability change that is noted for thresholds spanning a factor of 10.*

- **Soot production pathways and species influence**

Finally, the sensitivity of the influence metric and backbone evolution in increasing pressures is studied. Because the influence metric aggregates contributions from all pathways connecting a given node to soot, its value increases as the threshold is lowered. Reducing the threshold increases the number of enumerated pathways, although each individual pathway contributes less. For influential hubs, these numerous weak pathways can still add appreciable value to their overall influence, provided that a sufficiently large number of such paths pass through the hub. As the

threshold decreases, the influence metric is expected to converge toward a stable value that reflects the true importance of each node in the original, unpruned graph. Figure S15 shows the change in influence as threshold values are reduced from $\tau = 10^{-5}$ to $\tau = 5 \times 10^{-6}$. As observed, the influence of most species across all spatial locations changes only marginally. The largest variations occur for smaller species such as acetylene, ethylene, ethenyl, and butadiene, with maximum changes of approximately 4%. This behavior arises because these small species participate in a large number of weak, low-contribution pathways that collectively add to their influence score. Importantly, the limited sensitivity of the influence metric for the majority of species indicates that the dominant carbon-flow structure is robust to moderate changes in threshold, particularly in the selected range.

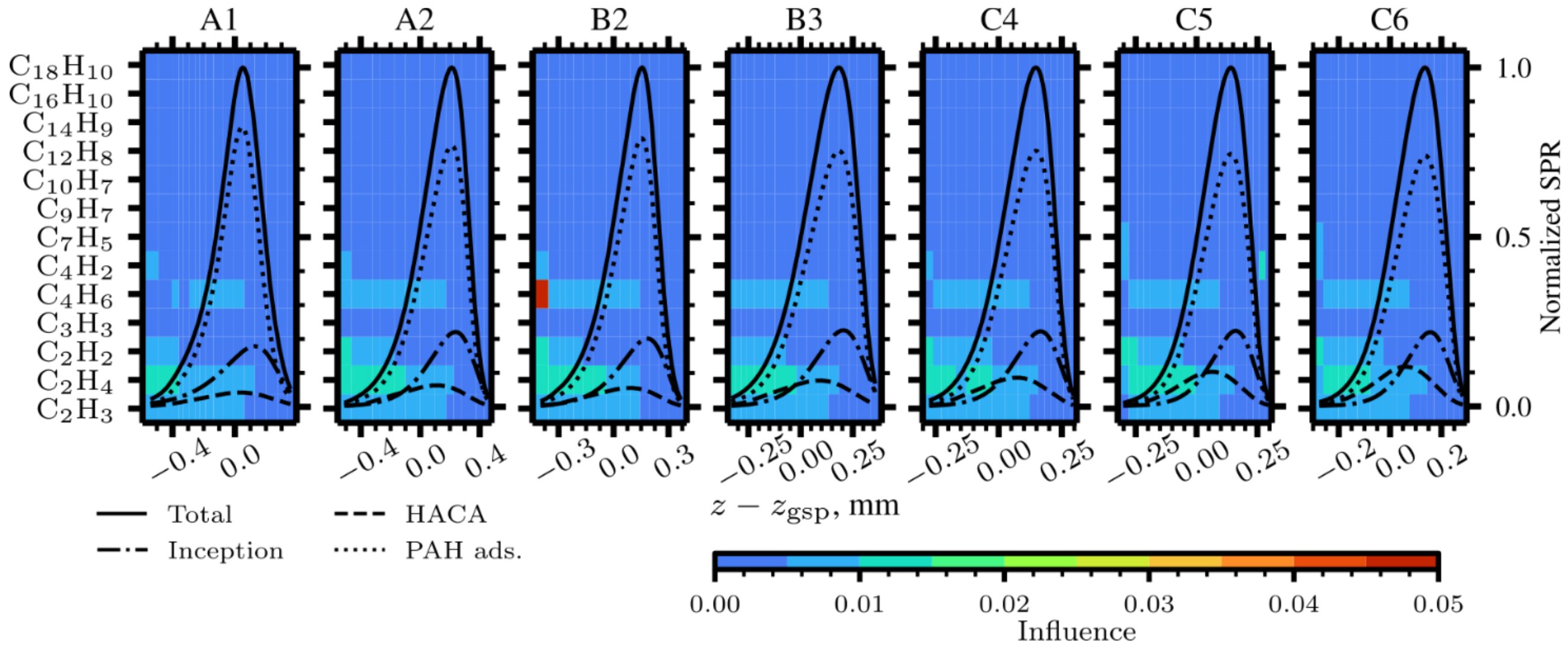


*Figure S15: Change in influence because of decreasing threshold from the selected $\tau = 10^{-5}$ to $\tau = 5 \times 10^{-6}$*

The backbone graphs and their change because of threshold variations are shown in Figure S16. As observed, the overall network structure remains largely unchanged, with only a small number of edges toggling on or off. These changes typically do not arise because the corresponding nodes or edges are entirely removed from the subgraph, but rather because they fall outside the top ten most influential nodes used to construct the backbone representation. In other words, the backbone differences reflect shifts in ranking rather than fundamental changes in connectivity.

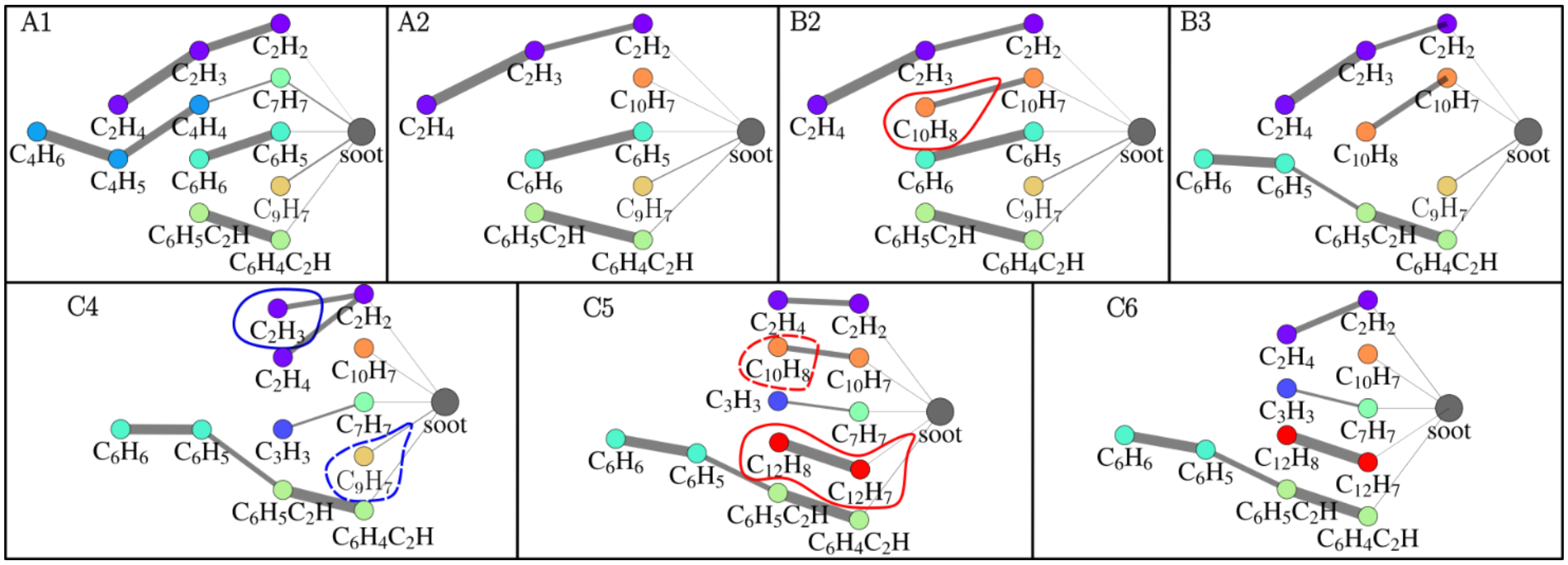


*Figure S16: Change in backbones through influential nodes with susceptible edges shown using blue and red circling. Solid and dashed circles show if the nodes are added or removed, respectively, while their colors show when they get added or removed (Blue shows the change when the threshold is decreased, while red shows the change when the threshold is increased)*


\fi

\newpage

\small
\baselineskip 10pt


\end{document}